%% file: main.tex
\documentclass[a4paper, 11pt]{article}

\usepackage[text={180mm,240mm},centering]{geometry}

\input{Preamble.tex}
\input{DefSymbols.tex}

\input{DefQuantities.tex}

\usepackage[maxbibnames=99]{biblatex}
\title{A nonlinear model of shearable  elastic rod from an origami-like  microstructure displaying  folding and faulting
}
\author[]{M. Paradiso}
\author[]{F. Dal Corso}
\author[]{D. Bigoni\footnote{Corresponding author: bigoni@ing.unitn.it}}
\affil[]{DICAM, University of Trento, via~Mesiano~77, I-38123 Trento, Italy}

\begin{document}
\date{Dedicated to Professor Pedro Ponte Casta\~neda on the occasion of his 65$^\text{th}$ birthday} 
\maketitle

\begin{abstract}
\noindent
A new continuous  model of {\it shearable} rod, subject to large elastic deformation, is derived from nonlinear homogenization of a one-dimensional periodic microstructured chain. As particular cases, the governing equations reduce to the Euler elastica and to the shearable elastica known as \lq Engesser', that has been scarcely analysed so far. 
The microstructure that is homogenized is made up of  elastic hinges and four-bar linkages, which may be realized in practice using origami joints. The equivalent continuous rod is governed by a Differential-Algebraic system of nonlinear Equations (DAE), containing an internal length ratio, and showing a surprisingly rich mechanical landscape, which involves a twin sequence of bifurcation loads, separated by a \lq transition' mode. The latter occurs, for simply supported and cantilever rods in a \lq bookshelf-like' mode and in a mode involving faulting (formation of a step in displacement), 
respectively. 
The postcritical response of the simply supported rod exhibits the emergence of folding, an infinite curvature occurring at a point of the rod axis, 
developing into a curvature jump at increasing load. Faulting and folding, excluded for 
both Euler and Reissner models and so far unknown in the rod theory, 
represent \lq signatures' revealing the origami 
design of the microstructure. These two 
features 
are shown to be associated
with bifurcations and, in particular folding, with a secondary bifurcation of the corresponding discrete chain when the number of elements is odd.
Beside the intrinsic theoretical relevance to the field of structural mechanics, our results can be applied to various technological contexts involving highly compliant mechanisms, such as the  achievement of objective trajectories with soft robot arms through folding and localized displacement of origami-inspired or multi-material mechanisms. 

\end{abstract}

\section{Introduction}

\begin{figure}[!h]
     \centering
        \includegraphics[width=180mm]{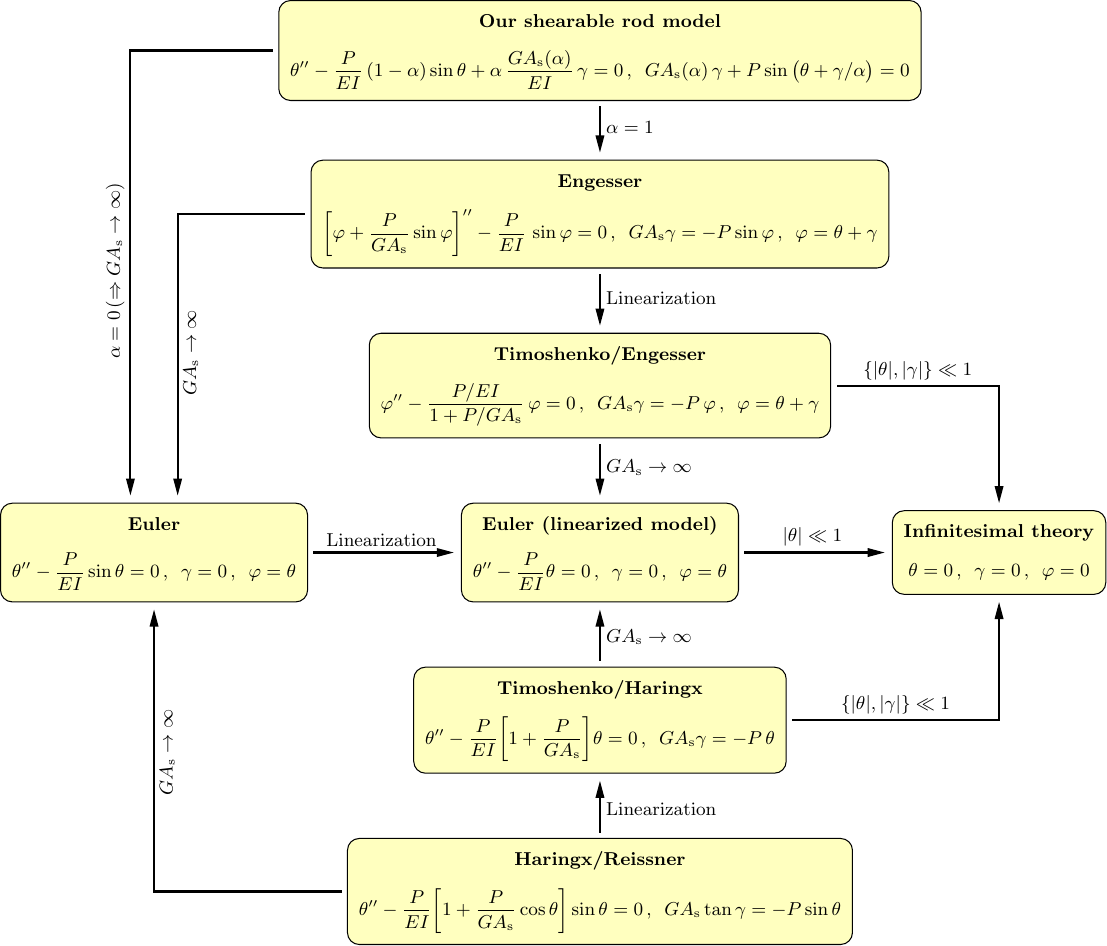}
     \caption{
     Different shearable models for a simply supported rod subject to an axial force $P$ (negative in compression).
     $\alpha$ is a micromechanical parameter, 
     $EI$ and $GA\ped{s}$ are, respectively, the bending and the shear stiffnesses of the rod's cross-section and 
     $\theta$ its rotation, different from the   rotation angle $\varphi$ of the tangent to the rod's axis, due to the presence of a shear angle $\gamma/\alpha$. The symbol $'$ denotes differentiation with respect to the axis coordinate $s$ (omitted for conciseness). Notice that: (i.) in the infinitesimal theory all the models are characterized by vanishing rotational fields for every value of the axial force $P$; (ii.) the rod model introduced here (labelled \lq our shearable rod model') is characterized by two coupled differential--algebraic equations, while in all other models the algebraic equation is decoupled from the differential one, which can be solved independently of the other involved variable.  
     }
     \label{fig:panel}
 \end{figure}

The implementation of shear compliance into an elastic rod model is an old mechanical problem,  rooted  in the early works of Rankine \cite{Rankine.1858},  Bresse \cite{Bresse.1859}, and Ehrenfest (following Elishakoff \cite{CHALLAMEL2019103389,Elishakoff.2020}), and Timoshenko \cite{Timoshenko.2020}). Though standard in the linear beam theory, the effect of shear becomes controversial when large deformation is involved, so that different approaches lead to different results and end up affecting even the linearized problem of buckling. This is a consequence of the fact that  
there are several ways to introduce both kinematics and internal forces in a large deformation setting and that, without violating any rule of mechanics, these ways reflect different points of view and lead to different results. 
So far, two main mechanical models for shear deformable rods have been pointed out, known as \lq Engesser' \cite{Engesser1891} and 
\lq Haringx' \cite{Haringx}, the latter formulated in full generality by Reissner \cite{Reissner1982}. The Reissner model is the mathematically easiest among the two and the most widespread 
\cite{attard, attard2, batista, cazzolliphd, simo},
while the former is only scarcely analysed \cite{atanackovic, Gjelsvik, Kocsis2018, ziegler}. 
Both the models reduce to:
(i.) the Euler's elastica when the shear deformation is neglected, and (ii.) to the two versions of the  prestressed Timoshenko model for the determination of beam buckling (both included by him in \cite{timoshenkostab}). When all terms are considered, Fig.~\ref{fig:panel} illustrates the hierarchy of models, with reference to a rod in a simply supported configuration with a load parallel to the initial straight axis.

The striking difference between the 
two Haringx/Reissner and the Engesser models can immediately be appreciated by considering that the former rod predicts bifurcation occurring under tensile force, which are impossible for the Engesser and Euler rods
(tensile bifurcations for Euler rod can be induced by constraints \cite{bigtens}).

Perhaps curiously, progress in clarifying shearable rod models can be made by resorting to prototyping technology (3D printing and CNC machines). In fact, the latter sets the designer free from the constraint of using homogeneous materials to realize a beam; rather, an elastic rod can be obtained through 
the periodic repetition of microstructures that may reach virtually any level of complexity.  
With a properly designed microstructured chain one can independently tune axial, shear, and bending stiffnesses to obtain desired compliance ratios.  
The overall behaviour of the chain can be homogenized into the response of an equivalent elastic rod. In the latter, the effective shear  and  bending stiffnesses are no longer mutually constrained as in the case of a rod made of a homogeneous material, so that different continuous models emerge and find their justification as connected to a given microstructure.  
Pioneers in this line of research, and extending earlier results of Domokos
\cite{Domokos1993,Domokos2002,Domokos1993109}, 
are Challamel and co-workers \cite{Challamel2021, Hache2018221, Kocsis2018, Kocsis, Lerbet20201}
who have shown how the Reissner and the Engesser models can  both be obtained by employing two microstructures differing only in the elastic element transmitting shear: a \lq Love finite shear strain' component 
(equivalent to a slider and similar to the element used in \cite{fraldi})
in the former case and a \lq Timoshenko finite shear strain' component in the latter. In this way, the two famous shearable rod models have been shown to be  both consistent from a mechanical perspective, but representative of two different microstructures. 

\begin{figure}[!h]
     \centering
        \includegraphics[width=120mm]{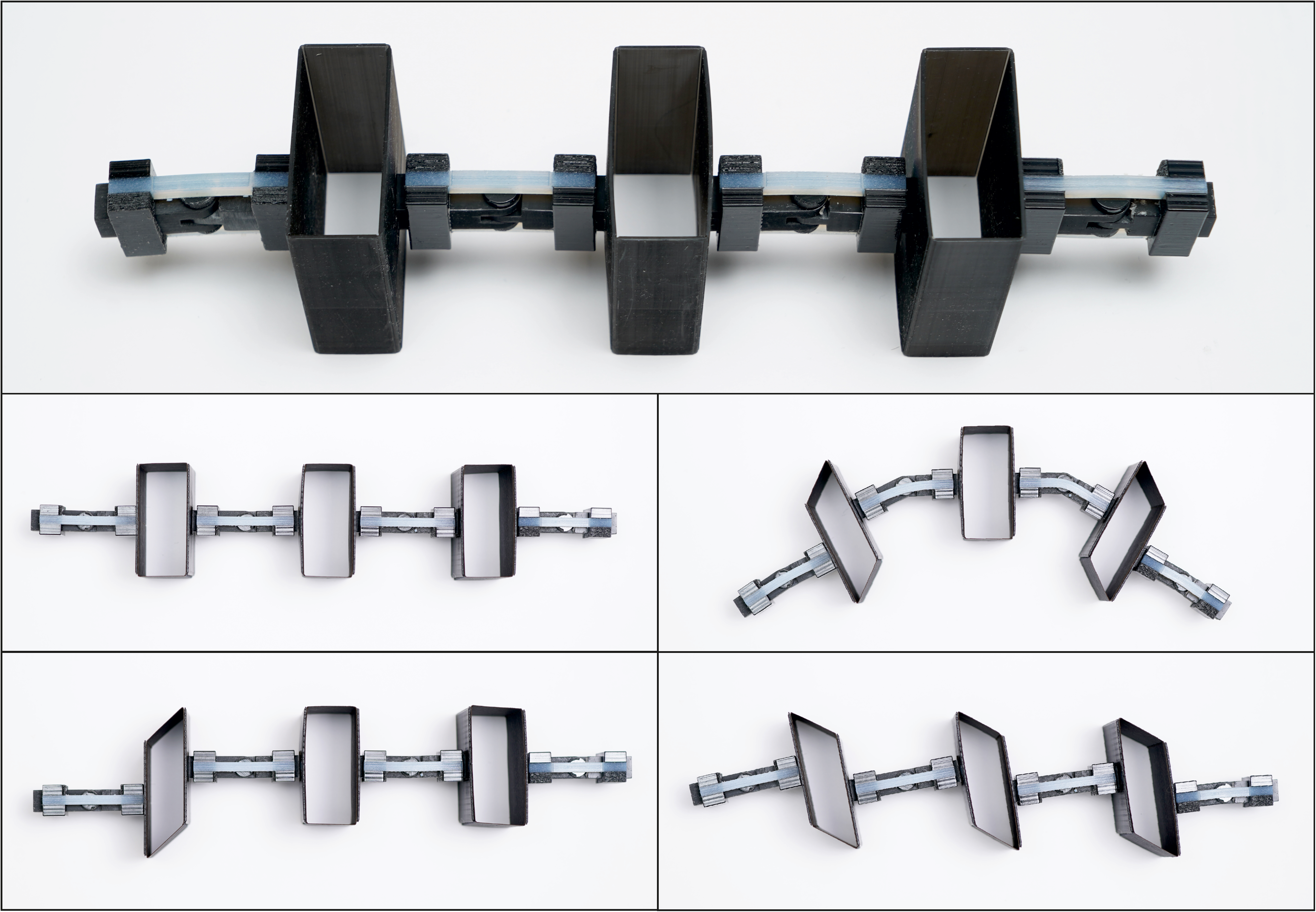}
     \caption{Concept prototype of the microstructure forming a three-element discrete chain to be homogenized into a new model of shear deformable elastic rod. Note the box-like elements representing origami-junctions and the white elements crossing the hinges, so providing the flexural stiffness. Upper part: lateral view of the undeformed configuration. Top views of the undeformed configuration (central part, left), of a bent deformation (central part, right), of the bifurcation modes  with the faulting (lower part, left) and the \lq bookshelf-like' (lower part, right) modes. 
     }
     \label{fig:intro}
 \end{figure}

The aim of the present article is to introduce, via homogenization, a new nonlinear model of shearable rod, based on a four-bar linkage microstructure, representative of a simple \lq origami' junction (a concept prototype of a three-element chain at macroscale\footnote{
     The black elements in the concept model are made in Loctite 3D IND403 and were manufactured with 
     a Stratasys Origin One 3D printer. They are jointed with folded cardboard sheets.  The white connections providing bending stiffness are made in Agilus30 and were 3D printed 
with a a Stratasys J750.
} 
is shown in Fig.~\ref{fig:intro}, 
but microscale realizations are possible\footnote{
Microscale compliant mechanisms, closely resembling the four-bar linkage system explored here, may be realised by means of advanced 3D-printing technologies based on 2PP (two-photon polymerization). Such  technologies, which are also referred to as direct laser writing, achieve sub-micron resolution. For instance, with such a 3D-printer it would be possible to manufacture a composite rod of total length 30~mm, containing 30 four-bar links, and cross section 0.5~mm$\times$0.5~mm in the thicker parts.}). 
In this model, 
an internal length ratio, 
constitutive and geometrical in nature and 
absent in the Engesser and Reissner rods, appears in the governing equations. 
The complexity of the microstructure leads to a work-conjugacy governing the decomposition of the force internal to the equivalent rod into a non-trivial definition of normal and shearing forces. The decomposition shows that the unshearable Euler and the shearable Engesser models can both be interpreted as limit cases of the broader shearable model introduced here, obtained for limit values of the internal length ratio, Fig.~\ref{fig:panel}. 
Our equivalent rod model is governed by a set of irreducible Differential--Algebraic Equations (DAE) and shows a rich mechanical behaviour. With respect to the Engesser model, our rod theory leads to a Sturm-Liouville problem for the bifurcation under compressive loads, 
surprisingly characterized 
by two \emph{twin} infinite sequences of eigenvalues. 
One sequence has eigenvalues bounded (in their absolute value) from both below and above, while the other sequence is bounded only from below. These two sequences are found to be separated by an additional  eigenvalue, which  corresponds to a \lq bookshelf-like' mode of bifurcation for a simply supported beam configuration, or to a discontinuous mode in a cantilever setting. 

For sufficiently high values of loading, our equivalent rod shows the emergence of discontinuous solutions in terms of either  folding (formation of a discontinuity in the curvature of the rod axis, in a simply supported scheme) or faulting (formation of a displacement jump at the clamp in a cantilever scheme). 
Such singular mechanisms in rod elements are analogous to the folding and faulting found in highly-anisotropic elastic solids, where the shear/bending stiffness ratio of the rod plays now a role analogous to the anisotropy contrast of the continuum \cite{gourgiotis2, gourgiotis1, gourgiotis3}.
These folding and faulting discontinuities 
are \lq inherited' by the presence of origami junctions in the microstructural design and 
were never so far observed in any rod model. While they are excluded for both Euler and Reissner rods, our model shows (as a particular case) that both discontinuities may arise for the Engesser rod, another previously overlooked feature. 
Folding initiates from a point 
(but we show also multiple folding, initiating simultaneously at two different cross sections,  associated to two points) 
along the rod axis where the curvature localizes, becoming infinite, and at the correspondent step of loading 
the Runge-Kutta integration  of the DAE governing the deformation of the rod fails to converge. The difficulty in the integration can be bypassed through the insertion at the singularity point in the equivalent rod of a special element localizing the curvature, namely, an elastic hinge.
Providing an  increasing stiffness to the hinge, it is shown that the solution converges to a constant result, evidencing folding.   
The latter is further explained through consideration of the 
microstructured chain underlying the homogenized response of the rod. 
In particular, when the chain is characterized by an even number of elements, the folding of the continuum simply becomes a localized rotation between the two central elements. The situation complicates when the number of elements  is odd, so that a secondary bifurcation is shown to occur, where the initial postcritical path becomes unstable and separates from a  new stable postcritical path, in which a sort of \lq cusp' of elements forms, mimicking the folding of the continuous model.

The results outlined in the present article introduce a new approach to the design of elastic structures by leveraging microstructural features and show that 
\begin{center}
\emph{many 
(or perhaps even infinite) 
new shearable rod models can be introduced, \\
with internal geometrical constraints (possibly different from our constraint involving an internal \\ length ratio), 
all representing real behaviours of periodic chains of discrete elements.}
\end{center}
This concept opens up a wide range of new applications, such as folding and faulting, which could be pivotal in developing soft robotic arms. Indeed, origami-inspired arms \cite{Kaufmann2022,Li2024,Novelino2020,Rus2018,Wu2021,Zhang2023} complement solutions based on soft continuous elements \cite{laschi,POLYGERINOS} and are currently defining new design strategies in soft robotics. Therefore, design breakthroughs can be guided by new structural concepts and models capable of displaying both global and local deformation mechanisms, which is the point of view pursued in the present article.




\section{Discrete one-dimensional system endowed with four-bar linkages}

\subsection{Geometry of the unit cell}
\begin{figure}[!h]
    \centering
       \includegraphics[width=170mm]{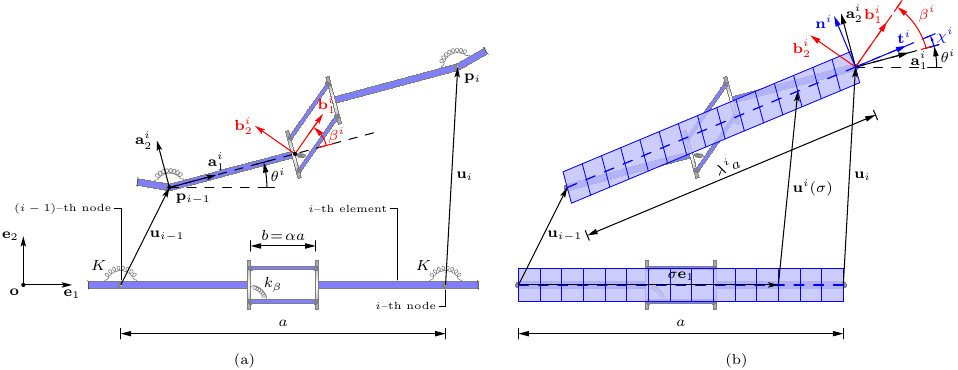}
    \caption{Undeformed (lower part) and deformed (upper part) configurations of the $i$--th unit cell (a) and the respective configurations of its homogenized counterpart (b).  The deformed state of the $i$--th unit cell, joining the $(i-1)$--th  and $i$--th nodes 
    (position vectors $\Bp_{i-1}$ and $\Bp_{i}$), is described by the two angles $\theta^i$ and $\beta^i$. 
    The piece-wise rigid motion of the cell is interpolated by a smooth displacement field for the homogenized $i$--th element. Its nonlinear elastic behaviour follows from the geometric 
    nonlinearity of the kinematics, involving relative rotations at the nodes (constrained by rotational springs of stiffness $K$) and relative displacements at the four-bar linkage (equipped with rotational springs of stiffness $k_\beta$). The unit vector $\Bt^i$ denotes the tangent (and $\Bn^i$ the normal) to the equivalent rod axis, while $\Ba^i_1$ is normal (and $\Ba^i_2$ tangent)  to the cross-section, while  
     $\Bb_1^i$ is aligned parallel (and $\Bb_2^i$ normal) to the linkage bars.
    }
    \label{fig:DiscreteLinkageKinem}
\end{figure}

Within a reference Cartesian  planar system $x_1$--$x_2$, defined by the two orthonormal vectors $\Be_1$ and $\Be_2$, the mechanical response is analysed for a one-dimensional structural system (of $n$ elements and $n+1$ nodes) realized as the periodic repetition of a unit cell of undeformed length $a$ aligned parallel to the direction $\mathbf{e}_1$, Fig.~\ref{fig:DiscreteLinkageKinem}a. The repetition of $n$ unit cells realizes a one-dimensional structure having a total length $L=n a$ and its undeformed axis aligned parallel to the direction $\mathbf{e}_1$. Symmetric with respect to both $\Be_1$ and $\Be_2$  in its undeformed state, the unit cell is  made up of two equal rigid end bars,  connected through a parallel four-bar linkage of length $b$ so that the ratio
\begin{equation}
    \label{lunghezzainterna}
    \alpha = \frac{b}{a} \in[0,1] \,, 
\end{equation}
becomes an internal length ratio, surviving the homogenization process and
thus affecting the nonlinear behaviour of the equivalent rod. The four-bar linkage forms a rectangle in the unloaded configuration and a parallelogram in any deformed state; it is equipped with a linear elastic rotational spring, of stiffness $k_\beta$.  
The $i$--th unit cell  is  joined to the $(i+1)$--th cell through a linear elastic hinge with rotational stiffness $K$, present at each internal node ($i=1,...,n-1$).

\subsection{Large deformation kinematics of the unit cell}

The current (deformed) configuration of the $i$--th unit cell  is completely described through its 4 degrees of freedom (d.o.f.), as it results from a rigid motion combined with the actuation of the four-bar mechanism.

To properly describe the motion of the $i$--th element ($i=1,...,n$), it is instrumental to introduce two local orthonormal bases: $\curlyb{\Ba_1^i,\Ba_2^i}$ and $\curlyb{\Bb_1^i,\Bb_2^i}$, respectively aligned and orthogonal to the rigid end bars and to the internal mechanism's bars, Fig.~\ref{fig:DiscreteLinkageKinem}a. The two orthonormal bases $\curlyb{\Ba_1^i,\Ba_2^i}$ and $\curlyb{\Bb_1^i,\Bb_2^i}$ are expressed 
for the $i$--th unit cell
as
\begin{equation}\label{eq:DiscLocalVectors}
    \Ba_j^i = \BR(\theta^i) \Be_j \,, \qquad \Bb_j^i = \BR(\theta^i + \beta^i)\Be_j \,, \quad i=1,...,n,\quad j = 1,2 \,,
\end{equation}
where $\theta^i$ and $\beta^i$ are respectively the rotation of the end   bars (equal to the overall rotation of the unit cell) and the relative rotation of the four-bar mechanism, while $\BR(\phi)$ is the rotation tensor associated to a rotation angle $\phi$
\begin{equation}\label{eq:RotationTensorExplicit}
\BR(\phi) = (\cos{\phi} \, \Be_1 + \sin{\phi} \, \Be_2) \otimes \Be_1 - (\sin{\phi} \, \Be_1 - \cos{\phi} \, \Be_2) \otimes \Be_2 \,,
\end{equation}
where the symbol $\otimes$ denotes the dyadic product and the rotations are considered positive when counter-clockwise.

The two end points of the $i$--th unit cell are defined by the two nodes  $\roundb{i-1}$ and $i$.
Introducing the position of the $i$--th node defined by vector $\Bp_i$, the relative position for the two nodes of the $i$--th unit cell is
\begin{equation}
    \Bp_i - \Bp_{i-1} = a\left[\roundb{1 - \alpha }\Ba_1^i + \alpha  \Bb_1^i \right]\,,
\end{equation}
or, equivalently from \cref{eq:DiscLocalVectors},
\begin{equation}\label{eq:DiscNodePosition}
    \Bp_i - \Bp_{i-1} = a  \squarec{\roundb{1-\alpha} \BR\roundb{\theta^i} + \alpha \BR\roundb{\theta^i + \beta^i}}\Be_1 \,.
\end{equation}

The kinematics of the internal linkage implies  a unilateral internal constraint on  the distance between two adjacent nodes, expressed as 
\begin{equation}\label{eq:DiscNodeDistance}
    \dfrac{\norma{\Bp_i - \Bp_{i-1}}}{a} = \sqrt{1 - 2 \alpha\roundb{1-\alpha}\roundb{1-\cos{\beta^i}}} \leq 1 \,,
\end{equation}
enforcing the distance between two nodes singling out a cell never to exceed   its initial value.
Moreover, \cref{eq:DiscNodePosition} can be written in terms of displacements as
\begin{equation}\label{eq:DiscNodeDisplacement}
    \Bu_i - \Bu_{i-1} = a  \squarec{\roundb{1-\alpha} \BR\roundb{\theta^i} + \alpha \BR\roundb{\theta^i + \beta^i}}\Be_1 - a\Be_1 \,,
\end{equation}
because the position of the $i$--th node in the reference (undeformed) configuration is $i\,a\Be_1$.

\subsection{Homogenization of the \texorpdfstring{$i$}{i}--th unit cell }

The presence of the four-bar linkage realizes a discontinuous  deformed shape of each unit cell. In order to  derive an equivalent continuous rod model, it is instrumental to  perform   an \lq intermediate' homogenization of the $i$--th  unit cell through a linear interpolation of the position and displacement fields inside of it, Fig.~\ref{fig:DiscreteLinkageKinem}b.
To this purpose, a local coordinate $\sigma \in \squareb{0,a}$ is introduced along each element, measuring the distance of the generic point along the axis of the $i$--th unit cell from the $(i-1)$--th node in the {\it reference} configuration. 

\subsubsection{Homogenized kinematics}

Through the linear interpolation between the two end nodes $i-1$ and $i$, \cref{eq:DiscNodePosition}, the point $\Bp^i(\sigma)$, describing the deformed axis of the cell, can be defined as 
\begin{equation}\label{eq:DiscFieldpp}
    \Bp^i(\sigma) = \Bp_{i-1} + \sigma \squarec{\roundb{1-\alpha} \BR\roundb{\theta^i} + \alpha \BR\roundb{\theta^i + \beta^i}}\Be_1 \,,
    \qquad
    \sigma \in [0,a] \,,
\end{equation}
and, using \cref{eq:DiscNodeDisplacement}, the displacement becomes
\begin{equation}\label{eq:DiscFielduu}
    \Bu^i(\sigma) = \Bu_{i-1} + \sigma \squarec{\roundb{1-\alpha} \BR\roundb{\theta^i} + \alpha \BR\roundb{\theta^i + \beta^i}}\Be_1 - \sigma \Be_1 \,,
    \qquad
    \sigma \in [0,a] \,,
\end{equation}
where the continuity of both the position and displacement fields at each node implies 
\begin{equation}
    \Bp^i(a) = \Bp^{i+1}(0) = \Bp_i\,,\qquad
    \Bu^i(a) = \Bu^{i+1}(0) = \Bu_i\,.
\end{equation}

The cross-section of the $i$--th unit cell, orthogonal to the  rod axis in the undeformed configuration, follows the rotation $\theta^i$ of the edge bars, Fig \ref{fig:DiscreteLinkageKinem}b. 
The tangent $\Bt^i$ and normal $\Bn^i$ unit vectors tangent to the 
axis of the 
$i$--th unit cell are 
\begin{equation}\label{eq:DiscAxisVectorsDef}
    \Bt^i = \BR\roundb{\chi^i} \Ba_1^i \,,
    \qquad
    \Bn^i = \BR\roundb{\chi^i} \Ba_2^i \,,
\end{equation}
where the \lq deviation angle' $\chi^i$  measures the angle between the normal $\Ba_1^i$ (tangent $\Ba_2^i$) to the cross-section of the unit cell  and the tangent $\Bt^i$ (normal  $\Bn^i$) to the axis of the unit cell, 
\begin{equation}\label{eq:DiscAxisAngleChi}
    \sin{\chi^i} =  \dfrac{\alpha\, \sin{\beta^i}}{\lambda^i} \,,
    \qquad
    \cos{\chi^i} =  \dfrac{\roundb{1-\alpha} + \alpha\, \cos{\beta^i}}{\lambda^i} \,.
\end{equation}

It is interesting to observe that  
the deviation angle $\chi^i$ is related to the angle $\beta^i$, defining the deformation of the four-bar linkage, and the angles 
introduce a shortening of the axis of the $i$--th unit cell. In fact, the kinematic constraint \eqref{eq:DiscNodeDistance} can be interpreted as  the longitudinal  stretch $\lambda^i$ of the $i$--th homogenized unit cell,  
\begin{equation}\label{eq:DiscAvegeStretch}
    \lambda^i = \sqrt{1 - 2 \alpha\roundb{1-\alpha}\roundb{1-\cos{\beta^i}}} \leq 1 \,.
\end{equation}

It is  finally noted that, alternatively to Eq. \eqref{eq:DiscAxisVectorsDef}, the pair of unit vectors $\{\Bt^i, \Bn^i\}$ can  be introduced as a linear combination of the pairs $\{\Ba_1^i,\Ba_2^i\}$ and $\{\Bb_1^i, \Bb_2^i \}$ as
\begin{equation}\label{eq:DiscAxisVectors}
    \Bt^i =  \dfrac{\roundb{1-\alpha} \Ba_1^i + \alpha\, \Bb_1^i}{\lambda^i} \,,
    \qquad
    \Bn^i =  \dfrac{\roundb{1-\alpha} \Ba_2^i + \alpha\, \Bb_2^i}{\lambda^i} \,.
\end{equation}

\subsubsection{External loads on the unit cell}

An external distributed load $\bar{\Bq}^i$, modelled as uniform for simplicity,  is considered to be applied along the $i$--th unit cell, 
while concentrated forces and moments are assumed acting at the ends of the cell. The load is assumed to be \emph{dead}, so that it is assigned on the reference configuration.

\begin{figure}[ht]
    \centering
    \includegraphics[width=170mm]{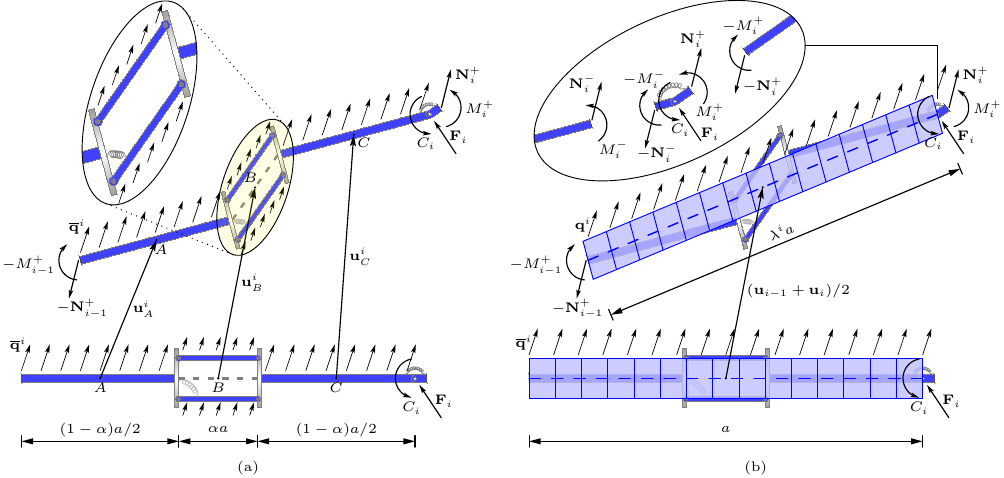}
    \caption{Dead load distribution (the external concentrated force $\BF_i$ and moment $C_i$ are applied at the $i$--th node, while the distributed load $\bar{\Bq}^i$ is defined in the reference configuration) and internal forces (transmitted by the contiguous elements from the left,  $\BN_{i}^-$ and $M_{i}^-$, and from the right, $\BN_{i}^+$ and $M_{i}^+$, of node $i$, see the inset) acting on the chain element 
    (a) and on the homogenized unit cell (b). 
    Part (a): $\Bu_A^i$, $\Bu_B^i$ and $\Bu_C^i$ denote displacements of the midpoints of the 3 rigid segments forming the $i$-th chain element;  the diffused load remains equal to $\bar{\Bq}^i$ when applied on the deformed configuration. Part (b): The equivalent load $\Bq^i =\bar{\Bq}^i/\lambda^i$ is distributed along the stretched axis of the homogenized unit cell, with average displacement $\roundb{\Bu_{i-i} + \Bu_i}/2$.
    }
    \label{fig:LinkageEquilibriumLoads}
\end{figure}

The homogenization of the uniform external load $\bar{\Bq}^i$ leads to a uniform external load $\Bq^i$ applied on the homogenized rod, equivalent in terms of work done to reach the deformed configuration. Note that the unit cell has a discontinuous shape, becoming continuous for its homogenized counterpart, Fig.~\ref{fig:LinkageEquilibriumLoads}. 

The load, acting on the length $\alpha a$ of the four-bar linkage, is divided in equal parts between the two longitudinal bars, Fig.~\ref{fig:LinkageEquilibriumLoads}a. Under this assumption, the work done by the dead load on the $i$--th element  can be evaluated by considering  the displacement of the midpoints, say, $A$, $B$ and $C$, of each of three connected rigid portions 
\begin{equation}
\begin{array}{l}
    \Bu^i_A =\Bu_{i-1} + \dfrac{a}{4} \roundb{1-\alpha}\squarec{\BR(\theta^i) - \BI}\Be_1  \,, \\ [2ex]
    \Bu^i_B =\Bu_{i-1} + \dfrac{a}{2} \squarec{\roundb{1-\alpha}\BR(\theta^i) + \alpha\BR\roundb{\theta^i + \beta^i} - \BI}\Be_1 \,, \\ [2ex]
    \Bu^i_C =\Bu_{i-1} + \dfrac{a}{4}\squarec{3\roundb{1-\alpha}\roundc{\BR(\theta^i) - \BI} + 4\alpha\roundc{\BR\roundb{\theta^i + \beta^i} - \BI}}\Be_1 \,.
\end{array}
\end{equation}
The work $\mathcal{W}^i\ped{discr}$ performed by $\bar{\Bq}^i$  can be written as follows 
\begin{equation}\label{eq:DiscLoadWorkSymm}
    \mathcal{W}^i\ped{discr} = a\,\bar{\Bq}^i \cdot \squared{\Bu_{i-1} 
        + \dfrac{a}{2}\roundd{\roundb{1-\alpha}\BR(\theta^i) + \alpha\BR\roundb{\theta^i + \beta^i} - \BI}\Be_1} \,.
\end{equation}

On the homogenized element, the load $\Bq_i$ has to be considered on the deformed configuration, 
so that a multiplication by the length of the  stretched axis, 
$a\lambda^i$, 
provides the resultant force. Moreover, the $i$--th displacement field is assumed linear, \cref{eq:DiscFielduu}, and thus the work of the external load can be obtained through a  multiplication by the average displacement $\roundb{\Bu_{i-1}+\Bu_i}/2$ as
\begin{equation}\label{eq:HomogLoadWorkSymm}
    \mathcal{W}^i\ped{homog} = a\, \lambda^i \Bq^i \cdot \squared{\Bu_{i-1} 
        + \dfrac{a}{2}\roundd{\roundb{1-\alpha}\BR(\theta^i) + \alpha\BR\roundb{\theta^i + \beta^i} - \BI}\Be_1} \,,
\end{equation}
where \cref{eq:DiscNodeDisplacement} has been used to evaluate the nodal displacement $\Bu_i$.

The equivalence between the works done on the discontinuous and on the homogenized element can be enforced by equating \cref{eq:DiscLoadWorkSymm} and \cref{eq:HomogLoadWorkSymm}. This equality is promptly obtained through a rescaling of
the load magnitude on the deformed configuration with the stretch $\lambda^i$, defined by \cref{eq:DiscAvegeStretch}, leading to 
\begin{equation}\label{eq:DiscDeadLoads}
    \lambda^i\Bq^i = \bar{\Bq}^i \,.
\end{equation}

Together with the load $\Bq^i$ distributed along the element, the dead concentrated force $\BF_i$ and couple $C_i$ are applied at the $i$--th node, the latter at its immediate left.
Considering the $i$--th element, connecting nodes $i-1$ and $i$, the unit cell is defined as the element joining the points at the immediate right of the nodes $i-1$ and $i$. In this way, the node $i$ and the concentrated loads acting on it are entirely included within the cell, while the node $i-1$ and its applied loads are excluded.

\subsubsection{Equilibrium and constitutive equations}

The equilibrium and constitutive relation for the $i$--th homogenized unit cell in its deformed state, subject to the equivalent distributed load $\Bq^i$, the  nodal force $\BF_i$ applied on the $i$-th node,  and the moment $C_i$ on the immediate left  of the $i$--th node, is ensured by the internal forces $\BN_{i-1}^+$ and $M_{i-1}^+$ applied at the right limit of the $(i-1)$--th node, as well as $\BN_{i}^+$ and $M_{i}^+$, acting at the right limit of the $i$--th node (see the inset of Fig.~\ref{fig:LinkageEquilibriumLoads}b).
The equilibrium of the $i$--th node is obtained by considering the left (labelled \lq $-$') and right (labelled \lq $+$') limits of the internal forces as 
\begin{equation}
\label{mazza}
   \BN_i^+- \BN^-_i+\BF_i = \Bnull \,,
   \qquad
    M_i^+-M^-_i+C_i = 0 \,.
\end{equation}

The dead loading is conservative, so that the equilibrium and constitutive relation of the unit cell can be analysed through the energy approach. The work $\mathcal{W}^i$ performed by the loads  on the $i$--th unit cell is given by the sum of the works done by the 
equivalent distributed load $\Bq^i$, \cref{eq:HomogLoadWorkSymm}, 
by the concentrated load $\BF_i$,  by the internal forces $\BN_{i-1}^+$ and $\BN_{i}^+$
(obtained through multiplication with the conjugate displacements $\Bu_{i-1}$ and $\Bu_i$), by  the concentrated moment $C_i$, and by the  internal moments $M_{i-1}^+$ and $M_{i}^+$
(obtained through multiplication with the conjugate rotations  $\theta^i$ and $\theta^{i+1}$). The expression for the nodal displacements difference,  \cref{eq:DiscNodeDisplacement}, allows the evaluation of the work $\mathcal{W}^i$ on the $i$-th unit cell as
\begin{equation}\label{eq:DiscCellWW}
\begin{aligned}
    \mathcal{W}^i = & \roundc{C_i-M_{i-1}^+}\theta^i + M_{i}^+\theta^{i+1} \,
    -
     \roundc{\BN_{i-1}^+ - \BN_{i}^+ - \BF_i - a \lambda^i \Bq^i} \cdot \Bu_{i-1} \\[1ex] 
     &+  \roundd{\BN_{i}^+ 
     + \BF_i + \dfrac{a}{2}\lambda^i\Bq^i} 
       \cdot a \squared{ \roundd{\roundb{1-\alpha} \BR\roundb{\theta^i} + \alpha \BR\roundb{\theta^i + \beta^i}}\Be_1-  \Be_1} \,.
\end{aligned}
\end{equation}

Note that, according to the definitions used in \cref{mazza}, the contributions of the internal forces $\BN_{i-1}^+$ and $M_{i-1}^+$ have been considered with the negative sign.

The elastic energy $\mathcal{E}^i$ associated with the $i$--th unit cell is stored within the linear rotational springs at the $i$--th node  (stiffness $K$)  and at the four-bar linkage (stiffness $k_\beta$), respectively through the  unit cell rotation difference $\roundb{\theta^{i+1}-\theta^i}$ and the linkage rotation  $\beta^i$, 
\begin{equation}\label{eq:DiscCellUU}
    \mathcal{E}^i = \dfrac{1}{2} K \left(\theta^{i+1} - \theta^i\right)^2 + \dfrac{1}{2} k_\beta {\beta^i}^2 \,.
\end{equation}

\cref{eq:DiscCellUU,eq:DiscCellWW} reveal that the kinematic parameters defining the status of the unit cell comprise also the rotation $\theta^{i+1}$. Therefore, these 5 parameters become: the nodal displacement vector $\Bu_{i-1}$, the linkage rotation $\beta^i$, and the unit cell rotations  $\theta^i$ and $\theta^{i+1}$.
Accordingly, the stationary of the total potential energy ($\mathcal{E}^i - \mathcal{W}^i$) with respect to such parameters leads to the equations
\begin{equation}\label{eq:DiscCellStatio}
\left\{
    \begin{array}{l}
    \BN_{i}^+ = \BN_{i-1}^+ - a\lambda^i \Bq^i - \BF_i \,, \\[2ex]
    k_\beta \beta^i = \BN^i(a/2) \cdot a \alpha  \Bb^i_2 \,, \\[2ex]
    M_i^+ = M_{i-1}^+ - \BN^i(a/2) \cdot a \lambda^i \Bn^i  - C_i\,, \\[2ex]
    K \roundb{\theta^{i+1} - \theta^i} = M_i^+ \,,
    \end{array}
\right.
\end{equation}
where $\BN^i(a/2)$ is the 
value at the central point of the unit cell  ($\sigma=a/2$) of the internal force $\BN^i(\sigma)$ 
\begin{equation}\label{eq:DiscFieldFF}
    \BN^i(\sigma) = \BN_{i-1}^+ - \sigma \lambda^i \Bq^i \,,
    \qquad
    \sigma \in [0,a] \,,
\end{equation}
while the unit vectors $\Bb_2^i$ and $\Bn^i$ in \cref{eq:DiscCellStatio}\ped{2,3} follow from the role played by the rotation angles $\theta^i$ and $\beta^i$ in \cref{eq:DiscCellWW}.
In fact, the derivative of the rotation tensor $\BR(\phi)$,  \cref{eq:RotationTensorExplicit}, with respect to the angle $\phi$ is $\BR(\phi)\BR(\pi/2)$, which used in \cref{eq:DiscCellWW} leads to
\begin{equation}
    \dfrac{\partial}{\partial \theta^i}\squarec{\BR\roundb{\theta^i} \Be_1} = \BR\roundb{\theta^i}\Be_2 = \Ba^i_2\,,
    \qquad
    \dfrac{\partial}{\partial \beta^i}\squarec{\BR\roundb{\theta^i+\beta^i} \Be_1} = \BR\roundb{\theta^i+\beta^i}\Be_2 = \Bb^i_2\,,
\end{equation}
whence the occurrence of the unit vectors $\Bb^i_2$ 
and 
(through 
\cref{eq:DiscAxisVectors}\ped{2}) 
$\Bn^i$ in \cref{eq:DiscCellStatio}.

The distributed load $\Bq^i$ can be expressed in terms of the equivalent load $\bar{\Bq}^i$, assigned in the undeformed configuration and therefore known, \cref{eq:DiscDeadLoads}. Moreover, an evaluation of $\lambda^i\Bn^i$ through \cref{eq:DiscAxisVectors} and the  use of Eqs.~\eqref{eq:DiscLocalVectors} to express the unit vectors $\Ba_2^i$ and $\Bb_2^i$, allow to express the internal force $\BN^i(\sigma)$, pertaining to the $i$--th unit cell, Eq.~\eqref{eq:DiscFieldFF}, as
\begin{equation}
\label{eq:DiscFieldForcesExplicit}
    \BN^i(\sigma) = \BN_{i-1}^+ - \sigma \bar{\Bq}^i \,,
    \qquad
    \sigma \in [0,a] \,,
\end{equation}
so that Eqs.~\eqref{eq:DiscCellStatio}, describing the equilibrium of the $i$--th unit cell, become
\begin{equation}\label{eq:DiscCellStatioExplicit}
\left\{
    \begin{array}{l}
    \BN_{i}^+ = \BN_{i-1}^+ - a \bar{\Bq}^i - \BF_i \,, \\[2ex]
    k_\beta \beta^i = -\alpha \roundd{a\BN_{i-1}^+ - \dfrac{a^2}{2} \bar{\Bq}^i} \cdot \squarec{{\sin\roundb{\theta^i + \beta^i}}\, \Be_1 - {\cos\roundb{\theta^i + \beta^i}}\, \Be_2 } \,, \\[2ex]
    M_i^+ = M_{i-1}^+ + \roundd{a\BN_{i-1}^+ - \dfrac{a^2}{2} \bar{\Bq}^i} \cdot \squarec{\roundb{1-\alpha}\roundb{ \sin{\theta^i}\, \Be_1 - \cos{\theta^i} \, \Be_2} \\[1ex]
    \hspace{35ex}
    + \alpha \roundc{\sin\roundb{\theta^i + \beta^i} \, \Be_1 - \cos\roundb{\theta^i + \beta^i} \, \Be_2 }}  - C_i\,, \\[1ex]
    K \roundb{\theta^{i+1} - \theta^i} = M_i^+ \,.
    \end{array}
\right.
\end{equation}

\subsection{Analysis of the periodic discrete structure 
}

It is recalled that  the chain is composed of $n$ elements, with $n + 1$ nodes, and its complete configuration depends on $2n + 2$ kinematic parameters, corresponding to the two components of the displacement vector $\Bu_0$, the $n$  bar rotations $\theta^i$, and the $n$ four-bar linkage  angles $\beta^i$ ($i = 1, \ldots, n$). At this stage, each element forming the rod can both be understood as a discrete mechanism of a chain or an equivalent element of a piece-wise continuous rod, as reported in  Fig.~\ref{fig:LinkageEquilibriumLoads}.

The equilibrium equations required to determine the configuration of the entire structure can be obtained by assembling the equations derived for the $i$--th  unit cell, with the first and the $n$--th contribution treated separately.
In particular, the first, $i=0$, and the last, $i=n$, node of the structure 
have to obey  equilibrium 
\begin{equation}\label{eq:DiscBoundEquil}
        \BN_0^+ = -\BF_0 \,, \qquad
        M_0^+ = -C_0 \,, \qquad
        \BN_n^- = \BF_n \,, \qquad
        M_n^- = C_n \,,
\end{equation}
in which the end forces $\BF_0$, $C_0$, $\BF_n$, and $C_n$ may either be  unknown reaction forces (conjugated to  constrained displacement components) or prescribed external loads. Consistently with the formalism adopted in  Fig.~\ref{fig:LinkageEquilibriumLoads}b, internal forces transmitted from the left neighbourhood on the first node, $i = 0$, are absent ($\BN_0^- = \Bnull$, $M_0^- = 0$); similarly,  internal forces are not transmitted from the right neighbourhood of the last node, $i = n$, ($\BN_n^+ = \Bnull$, $M_n^+ = 0$).

In order to simplify the notation, the resultant external force $\BQ^i$ is introduced as the sum of the concentrated and the distributed loads applied  
between the midpoint of the $i$--th element and the final node of the structure, excluding the boundary force $\BF_n$ (becoming a boundary condition)
\begin{equation}\label{eq:DiscGeneralForceLoad}
    \BQ^i = \sum_{k=i}^{n-1} \BF_k + \sum_{k=i}^n a \lambda^k \Bq^k -\dfrac{1}{2} a \lambda^i \Bq^i  \,, \quad i = 1, \ldots, n \,,
\end{equation}
which is related via translational equilibrium to the $i$--th internal force acting at $\sigma = a/2$, defined by \cref{eq:DiscFieldFF},  through the following relation 
\begin{equation}\label{eq:DiscGeneralForceInternal}
    \BN^i(a/2) = \BQ^i + \BF_n \,, \quad i = 1, \ldots, n \,.
\end{equation}

\cref{eq:DiscGeneralForceInternal} can be exploited in Eqs.~\eqref{eq:DiscCellStatio} to explicitly relate the equilibrium of the $i$--th unit cell to the external loads. Moreover, 
the equations for all cells can be assembled
to obtain the equilibrium equations for the complete structure in terms of external loads, end forces, and configurational parameters 
\begin{equation}\label{eq:DiscRodEquil}
\left\{
    \begin{array}{l}
    \BQ^1 + \dfrac{1}{2} a \lambda^1 \Bq^1 + \BF_0 + \BF_n = \Bnull \,, \\[2ex]
    k_\beta \beta^i - \roundb{\BQ^i + \BF_n} \cdot a \alpha \Bb^i_2 = 0 \,,
        \quad i = 1, \ldots, n \,, \\[2ex]
    K\roundb{\theta^1 - \theta^2} - \roundb{\BQ^1 + \BF_n} \cdot a \lambda^1 \Bn^1  - C_1 - C_{0} = 0 \,, \\[2ex]
    K \roundb{-\theta^{i-1} + 2 \theta^i - \theta^{i+1}} - \roundb{\BQ^i + \BF_n} \cdot a \lambda^i \Bn^i - C_i = 0 \,, 
        \quad i = 2, \ldots, n-1 \,, \\[2ex]
    K\roundb{\theta^{n} - \theta^{n-1}} - \roundb{\BQ^n + \BF_n} \cdot a \lambda^n \Bn^n - C_n = 0 \,. 
    \end{array}
\right.
\end{equation}

Eq.~\eqref{eq:DiscRodEquil}\ped{1}, expressing the overall translational  equilibrium of the chain, can be obtained by summing up \cref{eq:DiscCellStatio}\ped{1} when the index $i$ ranges between 1 and $n$, with the assumptions $\BN_0^+ = -\BF_0$ and $\BN_n^+ = \Bnull$. Eqs.~\eqref{eq:DiscRodEquil}\ped{2} can be derived by applying \cref{eq:DiscCellStatio}\ped{2}, written in terms of external loads, to all $n$ elements.
Eq.~\eqref{eq:DiscRodEquil}\ped{3} is the specialization of Eq.~\eqref{eq:DiscCellStatio}\ped{3} for $i = 1$, along with $M_0^+ = - C_0$ and using \cref{eq:DiscCellStatio}\ped{4} to evaluate $M_1^+$.
Eq.~\eqref{eq:DiscRodEquil}\ped{4} is obtained by using Eqs.~(\ref{eq:DiscCellStatio})\ped{3} and (\ref{eq:DiscCellStatio})\ped{4}.
Finally, Eq.~\eqref{eq:DiscRodEquil}\ped{5} can be derived setting $i = n$ in \cref{eq:DiscCellStatio}\ped{3}, with  $M_n^+ = 0$ and evaluating $M_{n-1}^+$ through \cref{eq:DiscCellStatio}\ped{4}.

Observe that the product $\lambda^i\Bq^i$ represents the distributed load $\bar{\Bq}^i$ expressed in the  undeformed configuration, \cref{eq:DiscDeadLoads}, whence the resultant force $\BQ^i$ of external loads, defined by \cref{eq:DiscGeneralForceLoad}, can be written as
\begin{equation}\label{eq:DiscGeneralForceLoadExplicit}
    \BQ^i = \sum_{k=i}^{n-1} \BF_k + \sum_{k=i}^n a \bar{\Bq}^k -\dfrac{1}{2} a \bar{\Bq}^i  \,, \quad i = 1, \ldots, n \,,
\end{equation}
and the solvability condition expressed in terms of overall translational equilibrium, \cref{eq:DiscRodEquil}\ped{1},  reads as
\begin{equation}\label{eq:DiscRodEquilFFExplicit}
    \sum_{k=1}^{n-1} \BF_k + \sum_{k=1}^n a \bar{\Bq}^k + \BF_0 + \BF_n = \Bnull \,.
\end{equation}
Using Eqs.~(\ref{eq:DiscAxisVectors}) and  \eqref{eq:DiscLocalVectors} to express $\lambda^i\Bn^i$ and the unit vectors  $\Ba_2^i$ and $\Bb_2^i$, respectively, Eqs.~\eqref{eq:DiscRodEquil}\ped{2}--\eqref{eq:DiscRodEquil}\ped{5} can be rearranged as
\begin{equation}\label{eq:DiscRodEquilExplicit}
\left\{
    \begin{array}{l}
    K\roundb{\theta^1 - \theta^2} + a \roundb{\BQ^1 + \BF_n} \cdot \squarec{\roundb{1-\alpha}\roundb{\sin{\theta^1}\, \Be_1 - \cos{\theta^1} \, \Be_2} \\[1ex]
    \hspace{21ex}
    + \alpha \roundc{\sin\roundb{\theta^1 + \beta^1} \, \Be_1 - \cos\roundb{\theta^1 + \beta^1} \, \Be_2}}
    - C_1 - C_{0} = 0 \,, \\[2ex]
    K \roundb{-\theta^{i-1} + 2 \theta^i - \theta^{i+1}}
    + a \roundb{\BQ^i + \BF_n} \cdot \squarec{\roundb{1-\alpha}\roundb{\sin{\theta^i}\, \Be_1 - \cos{\theta^i} \, \Be_2} \\[1ex]
    \hspace{28ex}
    + \alpha \roundc{\sin\roundb{\theta^i + \beta^i} \, \Be_1 - \cos\roundb{\theta^i + \beta^i} \, \Be_2}} - C_i = 0 \,,
    \quad i = 2, \ldots, n-1 \,, \\[2ex]
    K\roundb{\theta^{n} - \theta^{n-1}} + a \roundb{\BQ^n + \BF_n} \cdot \squarec{\roundb{1-\alpha}\roundb{\sin{\theta^n}\, \Be_1 - \cos{\theta^n} \, \Be_2} \\[1ex]
    \hspace{26ex}
    + \alpha \roundc{\sin\roundb{\theta^n + \beta^n} \, \Be_1 - \cos\roundb{\theta^n + \beta^n} \, \Be_2}} - C_n = 0 \,, \\[2ex]
    k_\beta \beta^i + a \roundb{\BQ^i + \BF_n} \cdot \alpha \squarec{{\sin\roundb{\theta^i + \beta^i}}\, \Be_1 - {\cos\roundb{\theta^i + \beta^i}}\, \Be_2 } = 0 \,,
        \quad i = 1, \ldots, n \,,
    \end{array}
\right.
\end{equation}
which is a system of $2n$ nonlinear algebraic equations for the unknowns  $\theta^i$ and $\beta^i$ governing the mechanics of the discrete chain. Depending on the constraints applied at the end nodes, it may happen that some components of the end forces ($\BF_0$, $C_0$, $\BF_n$, $C_n$) may become additional unknowns, but in this case the dual displacements are constrained and these constraints are to be added to Eqs.~\eqref{eq:DiscRodEquilExplicit}.

The kinematic parameters $\theta^i$ and $\beta^i$, obtained as solutions of Eqs.~\eqref{eq:DiscRodEquilExplicit}, can be used to evaluate the internal forces transferred at each node of the discrete chain. In particular, the moment transferred at the $i$--th elastic hinge can be obtained from \cref{eq:DiscCellStatioExplicit}\ped{4}, while the internal force $\BN_i^+$ can be evaluated by iteratively applying \cref{eq:DiscCellStatioExplicit}\ped{1}, together with the equilibrium condition at the first node, $\BN_0^+ = - \BF_0$, and summing all the contributions up the $i$--th node, leading to
\begin{equation}
\BN_i^+ = - \displaystyle\sum_{k=1}^i \roundc{a \bar{\Bq}^k + \BF_k} -\BF_0 \,.
\end{equation}

Moreover, using \cref{eq:DiscNodeDisplacement} and summing all the contributions up to the $i$--th node, the horizontal and vertical components of the nodal displacements can be expressed as
\begin{equation}\label{eq:DiscRodDisplacementExplicit}
\begin{array}{l}
\Bu_i\cdot\Be_1 = \Bu_0\cdot\Be_1 + \displaystyle\sum_{k=1}^i a\squarec{\roundb{1-\alpha}\cos{\theta^k} + \alpha \cos\roundb{\theta^k + \beta^k} - 1} \,,
\\[1ex]
\Bu_i\cdot\Be_2 = \Bu_0\cdot\Be_2 + \displaystyle\sum_{k=1}^i a\squarec{\roundb{1-\alpha}\sin{\theta^k} + \alpha \sin\roundb{\theta^k + \beta^k}} \,.
\end{array}
\end{equation}

Note that, together with \cref{eq:DiscRodEquilFFExplicit}, the solvability condition expressing the overall rotational equilibrium must be satisfied, but 
this condition is already 
implicitly contained in  Eqs.~\eqref{eq:DiscRodEquilExplicit}\ped{1}--\eqref{eq:DiscRodEquilExplicit}\ped{3}, 
because their sum 
\begin{equation}\label{eq:DiscRodEquilMMExplicit}
\displaystyle\sum_{i=1}^n a \roundb{\BQ^i + \BF_n} \cdot \squarec{\roundb{1-\alpha}\roundb{\sin{\theta^i}\, \Be_1 - \cos{\theta^i} \, \Be_2} + \alpha \roundc{\sin\roundb{\theta^i + \beta^i} \, \Be_1 - \cos\roundb{\theta^i + \beta^i} \, \Be_2}} 
-\displaystyle\sum_{i=1}^{n-1} C_i  - C_0 - C_n= 0 \,,
\end{equation}
represents the moment equilibrium.

\section{Homogenization of the chain into an equivalent elastic rod}

\begin{figure}[t]
    \centering
    \includegraphics[width=170mm]{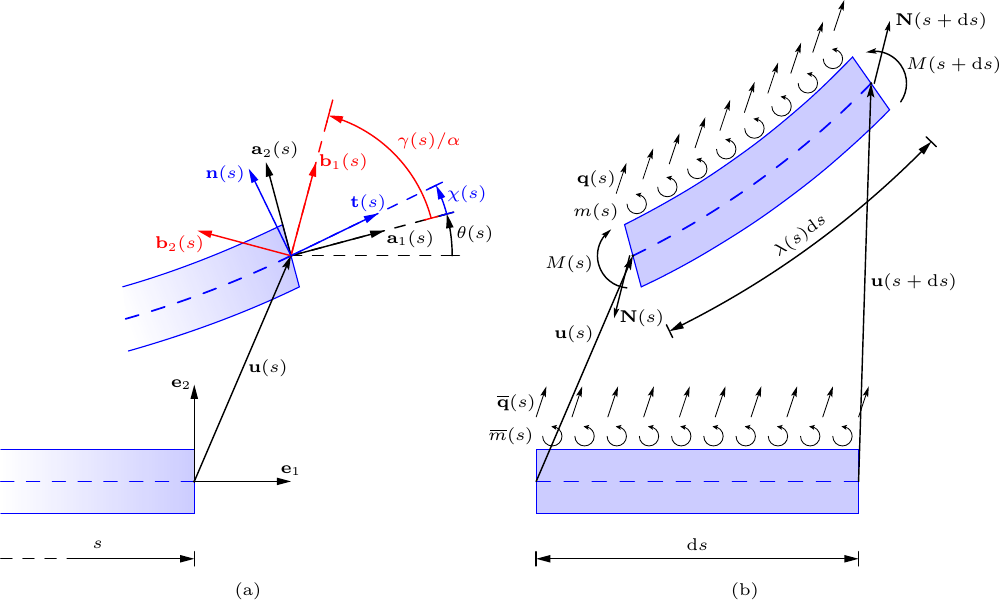}
    \caption{Deformation of the equivalent rod at the coordinate $s$ (a) and equilibrium of its infinitesimal  element (b).
    Part (a): Note the tangent $\Bt(s)$ and normal $\Bn(s)$ to the rod axis, the parallel $\Ba_2(s)$ and the normal $\Ba_1(s)$ to the rod cross-section, the inclination angle $\varphi(s) = \theta(s) + \chi(s)$ of $\Bt(s)$ and the shear angle $\gamma(s)/\alpha$. Part (b): Note the displacement $\Bu(s)$, the stretch $\lambda(s)$, the external ($\Bq(s)$ and $m(s)$) and the internal forces ($\BN(s)$ and $M(s)$). The internal force $\BN(s)$ is decomposed along the unit vectors $\Bb_1(s)$ and $\Bb_2(s)$, allowing an orthogonality in the sense of work.
    }
    \label{fig:LinkageContElem}
\end{figure}

The objective of this Section is to move from the equations governing the mechanics of 
the discrete chain to the differential equations ruling the behaviour of an equivalent elastic rod. 

\subsection{Kinematics and equilibrium of the infinitesimal rod element}\label{sec:ContinuousIndefElement}

The chain has an undeformed length $L = n a$, so that to homogenize its mechanical response, the length $L$ is kept fixed, while the number $n$ of unit cells is 
made to approach infinity ($n\to\infty$), so that  their length $a=L/n$ tends to zero. 
The equivalent rod is characterized by the coordinate $s \in [0,L]$, defined in the straight reference configuration, so that its infinitesimal element, comprised between $s$ and $s + \ds$, Fig.~\ref{fig:LinkageContElem}b, can be identified with the $i$--th unit cell shown in  Fig.~\ref{fig:LinkageEquilibriumLoads}b 
and the rotations defining the configuration of the element localize into a continuous model as
\begin{equation}\label{eq:ContRotationsLimit}
    \theta(s) = \lim_{a \to 0} \theta^i \,,
    \qquad
    \gamma(s) = \alpha  \lim_{a \to 0} \beta^i \,,
\end{equation}
where $\theta(s)$ becomes the continuously varying rotation of the cross-section of the rod, while $\gamma(s)$ is the cross-section kinematic parameter associated with the shearing slip. The multiplication by $\alpha$ in  the localization of $\beta^i$ to obtain  $\gamma(s)$  enforces 
the work-conjugacy of the internal forces with the displacement of the rod axis. 
In fact, applying \cref{eq:DiscNodeDisplacement} to small variation $\delta\beta^i$, with $\theta^i$ held constant, the variation of the relative displacement reads
\begin{equation}
\delta\roundb{\Bu_i - \Bu_{i-1}} = 
    a \BR\roundb{\theta^i+\beta^i} \Be_2 \, \alpha\, \delta\beta^i \,,
\end{equation}
and, through a localization of the internal force $\BN_{i-1}^+$ into its distributed version $\BN(s)$, the relevant work per unit length becomes
\begin{equation}
    \lim_{a \to 0} \BN_{i-1}^+ \cdot \BR\roundb{\theta^i+\beta^i} \Be_2 \, \alpha\, \delta\beta^i
    = \BN(s) \cdot  \BR\roundb{\theta(s) +\gamma(s)/\alpha} \Be_2 \, \delta\gamma \,.
\end{equation}

It should be observed that the angles $\theta(s)$ and $\gamma(s)/\alpha$ identify two distinct reference systems to be associated to the axis coordinate $s$, Fig.~\ref{fig:LinkageContElem}a. Specifically, a rotation of an angle $\theta(s)$ of vectors $\Be_j$ yields vectors  $\Ba_j$, while a further rotation of an angle $\gamma(s)/\alpha$ provides $\Bb_j$, representing the continuous counterparts of \cref{eq:DiscLocalVectors},
\begin{equation}\label{eq:ContSectVectors}
    \Ba_j(s) = \BR\roundb{\theta(s)} \Be_j \,, \qquad \Bb_j(s) = \BR\roundb{\theta(s) +\gamma(s)/\alpha}\Be_j \,, \quad j = 1,2 \,.
\end{equation}
On the other hand, the unit vectors tangent $\Bt(s)$ and normal  $\Bn(s)$ to the rod axis  are 
\begin{equation}\label{eq:ContAxisVectors}
    \Bt(s) =  \dfrac{\roundb{1-\alpha} \Ba_1(s) + \alpha\, \Bb_1(s)}{\lambda(s)} \,,
    \qquad
    \Bn(s) =  \dfrac{\roundb{1-\alpha} \Ba_2(s) + \alpha\, \Bb_2(s)}{\lambda(s)} \,,
\end{equation}
where $\lambda(s)$ is the stretch of the rod axis, intended as the reduction to a continuous model of the $i$--th stretch $\lambda^i$, \cref{eq:DiscAvegeStretch}, 
\begin{equation}\label{eq:ContStretch}
     \lambda(s) = \sqrt{1 - 2 \alpha\roundb{1-\alpha}\roundc{1-\cos\roundb{\gamma(s)/\alpha}}} \leq 1 \,.
\end{equation}

The unit vectors $\Bt(s)$ and $\Bn(s)$ represent the continualization of $\Bt^i$ and $\Bn^i$, respectively, defined by \cref{eq:DiscAxisVectorsDef}, so that the tangent to the rod axis deviates of an angle $\chi(s)$ with respect to the direction normal to the cross-section, Fig.~\ref{fig:LinkageContElem}a. Taking into account Eqs.\,(\ref{eq:DiscAxisAngleChi}) and (\ref{eq:ContRotationsLimit}), the deviation angle $\chi(s)$ satisfies
\begin{equation}\label{eq:ContAxisDeviationAngle}
    \sin{\chi(s)} =  \dfrac{\alpha \sin\roundb{\gamma(s)/\alpha}}{\lambda(s)} \,,
    \qquad
    \cos{\chi(s)} =  \dfrac{\roundb{1-\alpha} + \alpha \cos\roundb{\gamma(s)/\alpha}}{\lambda(s)} \,,
\end{equation}
and the sum of $\chi(s)$ with $\theta(s)$ provides the rotation angle of the tangent vector to the rod axis with respect to the horizontal direction, Fig.~\ref{fig:LinkageContElem}a, 
\begin{equation}\label{eq:ContAxisRotationAngle}
    \varphi(s) = \theta(s) + \chi(s) \,.
\end{equation}

More specifically, the ratio  $\roundb{\Bu_{i} - \Bu_{i-1}}/{a}$, \cref{eq:DiscNodeDisplacement}, in the limit of $n\to\infty$  becomes the derivative of the displacement 
 with respect to $s$, denoted  through a prime $(\;)'$ as
\begin{equation}\label{eq:ContElemDisp}
    \Bu'(s) = \squarec{\roundb{1-\alpha} \BR\roundb{\theta(s)} + \alpha \BR\roundb{\theta(s) +\gamma(s)/\alpha}}\Be_1 - \Be_1 \,,
\end{equation}
which defines the kinematics of the continuous rod model. The derivative of the rotation angle $\varphi(s)$, \cref{eq:ContAxisRotationAngle}, provides the axis curvature $\varphi'(s)$ as 
\begin{equation}\label{eq:ContAxisCurvature}
    \varphi'(s) = \theta'(s) +  \dfrac{\alpha + \roundb{1-\alpha} \cos\roundb{\gamma(s)/\alpha}}{\lambda^2(s)} \gamma'(s) \,,
\end{equation}
where Eqs.~\eqref{eq:ContAxisDeviationAngle} have been used together  with \cref{eq:ContStretch}.

The vanishing of the unit cell length $a$ shows that  $\BN^i(\sigma)$ converges to $\BN_{i-1}^+$ for any $\sigma$ in $[0,a]$, \cref{eq:DiscFieldFF}, a result that motivates the continualization of $\BN^i(a/2)$ as $\BN(s)$. Therefore, the continualization of Eqs.~\eqref{eq:DiscCellStatio} leads to the following homogenized equilibrium  and  constitutive equations for the equivalent rod 
\begin{equation}\label{eq:ContElemStatio}
\left\{
    \begin{array}{l}
    \BN'(s) + \lambda(s) \Bq(s) = \Bnull \,, \\[2ex]
    \dfrac{k_\beta}{a \alpha^2} \gamma(s) = \BN(s) \cdot \Bb_2(s) \,, \\[2ex]
    M'(s) + \lambda(s) \squareb{\BN(s) \cdot \Bn(s) + m(s)} = 0 \,, \\[2ex]
    aK \theta'(s) = M(s) \,.
    \end{array}
\right.
\end{equation}
The equilibrium equations ~\eqref{eq:ContElemStatio}\ped{1} and \eqref{eq:ContElemStatio}\ped{3} can be  obtained by evaluating the ratios $\roundb{\BN_{i}^+ - \BN_{i-1}^+}/{a}$ and $\roundb{M_{i}^+ - M_{i-1}^+}/a$ from Eqs.~\eqref{eq:DiscCellStatio}\ped{1} and \eqref{eq:DiscCellStatio}\ped{3}, respectively, in  the limit $a \to 0$. The equivalence between the loads distributed along the rod and those applied on the discrete chain stems from
\begin{equation}\label{eq:ContLoadLimit}
    \Bq(s) = \lim_{a \to 0} \roundd{\Bq^i + \dfrac{\BF_i}{\lambda^i a}} \,, \qquad  m(s) = \lim_{a \to 0} \roundd{\dfrac{C_i}{\lambda^i a}} \,.
\end{equation}

The above identification is based on the assumption that $\BF_i$ and $C_i$ remain in the limit  proportional to the length $\lambda^i a$, so that both limits of $\BF_i/(\lambda^i a)$ and $C_i/(\lambda^i a)$ converge. If not, the contributions of $\BF_i$ and $C_i$ cannot be smeared and must be modelled as concentrated forces and couples (a situation not addressed here for conciseness). In the case when the limit converges, the discrete forces $\BF_i$ and $C_i$, except $\BF_0$, $C_0$, $\BF_n$ and $C_n$, contribute to $\Bq$ and $m$ as divided by their length of influence. The forces $\BF_0$, $C_0$, $\BF_n$ and $C_n$ become end forces for the equivalent rod.

The constitutive 
relations, Eqs.\,\eqref{eq:ContElemStatio}\ped{4} and  \eqref{eq:ContElemStatio}\ped{2} respectively for the bending moment $M(s)$ and the shear force $\BN(s) \cdot \Bb_2(s)$, 
are obtained from Eqs.~\eqref{eq:DiscCellStatio}\ped{4} and \eqref{eq:DiscCellStatio}\ped{2}, taking into account Eqs.~\eqref{eq:ContRotationsLimit}. Notice that the former continualization also requires the limit \eqref{eq:ContLoadLimit}\ped{2} to be valid, so that $M_{i}^+$ approaches $M_{i-1}^+$ as the unit cell is contracted to null length, and can therefore be continualized as $M(s)$. 
 $\theta'(s)$ in Eq.~\eqref{eq:ContElemStatio}\ped{4} results in the limit of the ratio $\roundb{\theta^{i+1}-\theta^i}/a$. It is further noted that 
 the axial force component $\BN(s) \cdot \Bb_1(s)$ has to be understood as a Lagrangian multiplier in the model, because the rod is axially constrained,  Eq.~\eqref{eq:ContStretch}.

It is worth to highlight that, even if the homogenization of the discrete chain is based on the limit in which the length $a$ approaches zero, 
the parameters $k_\beta/(a\alpha^2)$ 
and $aK$, Eqs.~\eqref{eq:ContElemStatio}\ped{2, 4}, 
are assumed to remain finite in this limit.
Therefore, the nomenclature 
\begin{equation}\label{eq:ContSectionStiffness}
    GA\ped{s} = \dfrac{k_\beta}{a \alpha^2} \,,
    \qquad
    EI = a K \,,
\end{equation}
is introduced to keep contact with the continuous theories of shearable rods, so that the two constants should not be intended as related through the usual constants pertaining to  elastic  rods made of homogeneous material, but become independent parameters connected to the stiffnesses of the microstructure.
Moreover, during the homogenization, 
the geometric ratio $\alpha$, \cref{lunghezzainterna}, does not disappear, but ends up playing the role of an internal dimensionless length.

Observe finally that, while \cref{eq:ContElemStatio}\ped{4} represents the typical linear bending relation between the internal moment $M(s)$ and the curvature $\theta'(s)$, \cref{eq:ContElemStatio}\ped{2} provides a shear constitutive law which is peculiar to the model under consideration. In fact, the parameter $\gamma(s)$ plays the role of a shear slip angle, linearly related to the shear force $T(s)$, which is obtained by projecting the internal force $\BN(s)$ along the direction of the unit vector $\Bb_2(s)$.

The system $\curlyb{\Bb_1,\Bb_2}$, identified with the shear angle $\gamma(s)/\alpha$, is different from both the reference systems $\{\Ba_1,\Ba_2\}$ and $\{\Bt,\Bn\}$, 
with $\Ba_1$ orthogonal to the cross-section and  $\Bt$ tangent to the rod axis, Fig.~\ref{fig:LinkageContElem}a. Therefore, it follows that:
\begin{center}\emph{
the present equivalent shearable rod model defines\\ a decomposition of the internal force through shear and normal components  \\ neither aligned with the tangent to the cross-section (as in the Haringx/Reissner model \cite{Reissner1982}) \\nor  with the normal to the rod axis (as in the Ziegler model \cite{ziegler}).}
\end{center}

Moreover, a-priori assumptions about the shear direction are not needed, because the constitutive relation \cref{eq:ContElemStatio}\ped{2} is derived by a continualization process. On the contrary, the shear direction is implicitly included in the model as a cross-section property reflecting the microstructural properties of the discrete chain.

Note that  Eqs.~\eqref{eq:ContLoadLimit} introduce the distributed loads $\Bq(s)$ and couples $m(s)$ with reference to the deformed configuration of the equivalent rod. However, all loads are assumed to be \emph{dead}, so that in analogy to the discrete unit cell, \cref{eq:DiscDeadLoads}, the distributed loads $\bar{\Bq}(s)$ and couples $\bar{m}(s)$ can be associated to the undeformed configuration of the rod as  
\begin{equation}\label{eq:ContDeadLoads}
    \bar{\Bq}(s) = \lambda(s) \Bq(s) \,, 
    \qquad
    \bar{m}(s) = \lambda(s) m(s) \,,
\end{equation}
obtained by scaling $\Bq(s)$ and $m(s)$ though the rod axis stretch $\lambda(s)$, \cref{eq:ContStretch}.

In view of Eqs.~\eqref{eq:ContDeadLoads}, and using \cref{eq:ContStretch} together with Eqs.~\eqref{eq:ContSectVectors} to express $\lambda(s) \Bn(s)$ and $\Bb_2(s)$, the equilibrium equations of the rod element, Eqs.~\eqref{eq:ContElemStatio}\ped{1} and \eqref{eq:ContElemStatio}\ped{3}, can be written as
\begin{equation}\label{eq:ContElemEquilExplicit}
\left\{
    \begin{array}{l}
    \BN'(s) + \bar{\Bq}(s) = \Bnull \,, \\[2ex]
    M'(s) - \BN(s) \cdot \squarec{\roundb{1-\alpha} \roundc{\sin{\theta(s)}\, \Be_1 - \cos{\theta(s)}\, \Be_2} \\[1ex]
    \hspace{16ex}
    + \alpha\, \roundc{\sin{\roundb{\theta(s) +\gamma(s)/\alpha}}\, \Be_1 - \cos{\roundb{\theta(s) +\gamma(s)/\alpha}}\, \Be_2}} + \bar{m}(s) = 0 \,,
    \end{array}
\right.
\end{equation}
and the rod constitutive relations, Eqs.~\eqref{eq:ContElemStatio}\ped{2} and \eqref{eq:ContElemStatio}\ped{4}, can be expressed as
\begin{equation}\label{eq:ContConstRelExplicit}
    GA\ped{s} \gamma(s) = - \BN(s) \cdot \roundc{\sin{\roundb{\theta(s) +\gamma(s)/\alpha}}\, \Be_1 - \cos{\roundb{\theta(s) +\gamma(s)/\alpha}}\, \Be_2} \,,
    \qquad
    EI\, \theta'(s) = M(s) \,,
\end{equation}
where the equivalent shear $GA\ped{s}$ and bending $EI$ stiffnesses, Eqs.~\eqref{eq:ContSectionStiffness}, have been used.

\subsection{The mechanics of the equivalent rod}\label{sec:ContinuousRod}

Analogously to the analysis of the discrete chain, it is useful to introduce the resultant of the external  diffused load, $\BQ(s)$, defined at $s$ as
\begin{equation}\label{eq:ContGeneralForceLoad}
    \BQ(s) = \int_s^{L} \lambda(\sigma) \Bq(\sigma) \,\dsigma \,,
\end{equation}
where $\lambda(\sigma)$ is the stretch of the rod axis, \cref{eq:ContStretch}, and $\Bq(\sigma)$ the distributed load in the deformed configuration, \cref{eq:ContLoadLimit}\ped{1}. The product $\lambda(\sigma)\dsigma$ accounts for the change of variable, the axis coordinate, from the deformed configuration, where the load $\Bq$ is defined, to the undeformed one. The integral in \cref{eq:ContGeneralForceLoad} is consistently defined between the material axis coordinate $s$ and the undeformed rod length $L$.
In the same vein, the product $\lambda(\sigma)\Bq(\sigma)$ represents the equivalent load referred to the undeformed configuration, \cref{eq:ContDeadLoads}\ped{1}, whence the resultant force $\BQ(s)$ can  conveniently be expressed as
\begin{equation}\label{eq:ContGeneralForceLoadMaterial}
    \BQ(s) = \int_s^{L} \bar{\Bq}(\sigma) \,\dsigma \,.
\end{equation}

At the ends $s=0$ and $s=L$, the forces internal to  the equivalent rod correspond to the nodal forces [($\BF_0$ and $C_0$) and ($\BF_L$ and $C_L$), respectively]
\begin{equation}\label{eq:ContBoundEquil}
        \BN(0) = -\BF_0 \,, \quad
        M(0) = -C_0 \,, \quad
        \BN(L) = \BF_L \,, \quad
        M(L) = C_L \,.
\end{equation}

\cref{eq:ContGeneralForceLoad} can be viewed as the continualization of \cref{eq:DiscGeneralForceLoad}, where the term $-a\lambda^i\Bq^i/2$ vanishes when $a$ approaches zero, while Eqs.~(\ref{eq:ContBoundEquil})  are the continuous counterpart of Eqs.~\eqref{eq:DiscBoundEquil}. Consequently, the continualization of Eqs.~\eqref{eq:DiscRodEquil} provides the equilibrium equations for  the equivalent rod in the following form
\begin{equation}\label{eq:ContRodEquil}
\left\{
\begin{array}{l}
    \BQ(0) + \BF_0 + \BF_L = \Bnull \,, \\[2ex]
    GA\ped{s}\, \gamma(s) - \roundc{\BQ(s) + \BF_L} \cdot \Bb_2(s) = 0 \,, \\[2ex]
    EI\, \theta'(0) + C_{0} = 0 \,, \\[2ex]
    EI\, \theta''(s) + \roundc{\BQ(s) + \BF_L} \cdot \lambda(s) \Bn(s) + \lambda(s) m(s) =0 \,, \\[2ex]
    EI\, \theta'(L) - C_L = 0 \,,
\end{array}
\right.
\end{equation}
where $\Bb_2(s)$ is the shear direction, \cref{eq:ContSectVectors}\ped{2}, and $\Bn(s)$ is the normal to the rod axis, \cref{eq:ContAxisVectors}\ped{2}.

It is straightforward to derive Eq.~\eqref{eq:ContRodEquil}\ped{1} from Eq.~\eqref{eq:DiscRodEquil}\ped{1}, because $\BQ^1$ approaches $\BQ(0)$ and $a \lambda^1 \Bq^1/2$ vanishes when $a$ vanishes. The continualization of Eq.~\eqref{eq:DiscRodEquil}\ped{2} into Eq.~\eqref{eq:ContRodEquil}\ped{2} requires a preliminary division through  $a \alpha$ and use of \cref{eq:ContRotationsLimit}\ped{2}, because the ratio $k_\beta/(a\alpha^2)$ represents, in the limit, the shear stiffness of the equivalent rod, \cref{eq:ContSectionStiffness}\ped{1}.

The continualization of Eqs.~\eqref{eq:DiscRodEquil}\ped{3}--\eqref{eq:DiscRodEquil}\ped{5} into Eqs.~\eqref{eq:ContRodEquil}\ped{3}--\eqref{eq:ContRodEquil}\ped{5} can be obtained by applying  \cref{eq:ContSectionStiffness}\ped{2}. Eqs.~\eqref{eq:ContRodEquil}\ped{3} and \eqref{eq:ContRodEquil}\ped{5} are obtained in the limit $a \to 0$, noting that the ratios $\roundb{\theta^2-\theta^1}/a$ and $\roundb{\theta^n-\theta^{n-1}}/a$ become $\theta'(0)$ and $\theta'(L)$, respectively, and the terms having $a$ as a coefficient vanish, while $C_0$ and $C_L$ remain as concentrated moments at the rod ends. 
To obtain Eq.~\eqref{eq:ContRodEquil}\ped{4}, 
a division through $a$ is needed so that $\theta''(s)$ is obtained in the limit of the ratio $\roundb{\theta^{i-1} - 2 \theta^i + \theta^{i-1}}/a^2$ at vanishing $a$. The term $\lambda(s)m(s)$ is obtained using \cref{eq:ContLoadLimit}\ped{2}.

Note that 
Eqs.~\eqref{eq:ContRodEquil} can be alternatively derived from Eqs.~\eqref{eq:ContElemStatio}. Moreover,  \cref{eq:ContRodEquil}\ped{1} represents the so-called \lq solvability condition', namely, the overall translational equilibrium, which can be explicitly expressed in terms of the distributed load $\bar{\Bq}$, assigned in the undeformed configuration, as
\begin{equation}\label{eq:ContRodEquilFFExplicit}
    \int_0^{L} \bar{\Bq}(\sigma) \,\dsigma  + \BF_0 + \BF_L = \Bnull \,,
\end{equation}
where \cref{eq:ContGeneralForceLoadMaterial} has been used to evaluate the resultant external force $\BQ(s)$ at $s = 0$.

Furthermore, using Eqs.~\eqref{eq:ContAxisVectors}\ped{2}  and \eqref{eq:ContSectVectors} to evaluate $\lambda(s)\Bn(s)$, and the unit vectors $\Ba_2(s)$ and $\Bb_2(s)$, respectively, Eqs.~\eqref{eq:ContRodEquil}\ped{2} and \eqref{eq:ContRodEquil}\ped{4} can be rearranged into

\begin{center}
\emph{
the nonlinear system of DAE governing the equilibrium  at large deformation  of the rod, equivalent to the discrete chain, Eqs.~(\ref{eq:DiscRodEquilExplicit}), 
subject to arbitrary diffused and concentrated loads
}
\end{center}
\begin{equation}\label{eq:ContRodEquilDAE}
\left\{
\begin{array}{l}
    EI\, \theta''(s) - \roundb{1-\alpha} \roundc{\BQ(s) + \BF_L} \cdot \roundc{\sin{\theta(s)}\, \Be_1 - \cos{\theta(s)}\, \Be_2} 
    + \alpha\,GA\ped{s}\, \gamma(s) + \bar{m}(s) =0 \,, \\[2ex]
    GA\ped{s}\, \gamma(s) + \roundc{\BQ(s) + \BF_L} \cdot \roundc{\sin{\roundb{\theta(s) +\gamma(s)/\alpha}}\, \Be_1 - \cos{\roundb{\theta(s) +\gamma(s)/\alpha}}\, \Be_2} = 0 \,,
\end{array}
\right.
\end{equation}
\begin{center}
\emph{ containing the bending and shear stiffnesses, $EI$ and $GA\ped{s}$,  Eqs.~(\ref{eq:ContSectionStiffness}), and the internal length $\alpha$, \cref{lunghezzainterna}. 
}
\end{center}
\vspace{1 mm}

Eqs.~\eqref{eq:ContRodEquilDAE} has to be supplemented by Eqs.~\eqref{eq:ContRodEquil}\ped{3} and \eqref{eq:ContRodEquil}\ped{5} providing the boundary conditions
\begin{equation}\label{eq:ContRodEquilBC}
    EI\, \theta'(0) = -  C_0 \,,
    \qquad
    EI\, \theta'(L) = C_L \,.
\end{equation}

Solving Eqs.~\eqref{eq:ContRodEquilDAE} provides the  cross-section rotation $\theta(s)$  and the shear angle $\gamma(s)$, from which the rod axis stretch $\lambda(s)$ can be evaluated through \cref{eq:ContStretch}. Moreover, the axis deviation angle $\chi(s)$ can be evaluated by solving Eqs.~\eqref{eq:ContAxisDeviationAngle} and the rotation angle of the tangent to the axis can be obtained as $\varphi(s) = \theta(s) + \chi(s)$, \cref{eq:ContAxisRotationAngle}, and the axis curvature $\varphi'(s)$ is obtained from \cref{eq:ContAxisCurvature}.

The bending moment distribution along the equivalent rod can be  evaluated from \cref{eq:ContConstRelExplicit}\ped{2}, while the internal force $\BN(s)$ requires an integration of \cref{eq:ContElemEquilExplicit}\ped{1},  which, taking into account the boundary condition at $s = 0$, \cref{eq:ContBoundEquil}\ped{1}, yields 
\begin{equation}\label{eq:ContInternalForceInteg}
    \BN(s) = -\BF_0 - \int_0^s \bar{\Bq}(\sigma) \,\dsigma \,.
\end{equation}

The normal and shear components of the internal force $\BN(s)$ are associated with the unit vectors $\Bb_1(s)$ and $\Bb_2(s)$, \cref{eq:ContAxisVectors}\ped{2}. Specifically, the shear component can be easily obtained from the shear angle $\gamma(s)$, by exploiting the constitutive relation \eqref{eq:ContElemStatio}\ped{2}, as
\begin{equation}\label{eq:ContInternalShearForce}
    \BN(s) \cdot \Bb_2(s) = GA\ped{s}\, \gamma(s) \,,
\end{equation}
with $GA\ped{s}$ being the cross-section shear stiffness of the equivalent rod, \cref{eq:ContSectionStiffness}\ped{1}. The \lq normal' (i.e. orthogonal to the shear) component of the internal force  $\BN(s)$ can be evaluated by projecting \cref{eq:ContInternalForceInteg} onto $\Bb_1(s)$, to obtain
\begin{equation}\label{eq:ContInternalNormalForce}
    \BN(s) \cdot \Bb_1(s) = - \roundc{\cos{\roundb{\theta(s) +\gamma(s)/\alpha}}\, \Be_1 + \sin{\roundb{\theta(s) +\gamma(s)/\alpha}}\, \Be_2} \cdot \squaree{\BF_0 + \int_0^s \bar{\Bq}(\sigma) \,\dsigma} \,.
\end{equation}

The angles $\theta(s)$ and $\gamma(s)$, obtained as solutions of the DAE \eqref{eq:ContRodEquilDAE}, can in turn be employed as known functions for the differential equation defining the rod axis displacement $\Bu(s)$, \cref{eq:ContElemDisp}, which in  components becomes\footnote{The displacement derivative components  $u_j'(s)$ satisfy
$\lambda(s)=\sqrt{\left(1+u_1'(s)\right)^2+\left(u_2'(s)\right)^2}$.}
\begin{equation}\label{eq:ContDispDE}
    u_1'(s) = \roundb{1-\alpha} \cos{\theta(s)} + \alpha \cos{\roundb{\theta(s) +\gamma(s)/\alpha}} - 1 \,,
    \qquad
    u_2'(s) = \roundb{1-\alpha} \sin{\theta(s)} + \alpha \sin{\roundb{\theta(s) +\gamma(s)/\alpha}} \,,
\end{equation}
to be complemented with the displacement boundary conditions at $s = 0$ and $s = L$.

Once the displacement field $\Bu(s)$ is determined from integration of Eqs.~\eqref{eq:ContDispDE}, the current position $\mathbf{x}(s,y)$ of the generic point within any cross-section of the rod axis, at the transverse coordinate $y$, can be evaluated as
\begin{equation}
    x_1(s,y)=s+u_1(s)- y \sin\theta(s) \,,
    \qquad
    x_2(s,y)=u_2(s)+ y \cos\theta(s) \,.
\end{equation}

In closure, it should be noted that an integration of \cref{eq:ContRodEquilDAE}\ped{1} from $s = 0$ to $s = L$, together with the boundary conditions, Eqs.~\eqref{eq:ContRodEquilBC}, results to be 
\begin{equation}\label{eq:ContRodEquilMMExplicit}
\begin{array}{l}
\displaystyle
     \int_0^L \squarec{\roundb{1-\alpha} \roundc{\BQ(s) + \BF_L} \cdot \roundc{\sin{\theta(s)}\, \Be_1 - \cos{\theta(s)}\, \Be_2} - \alpha\,GA\ped{s}\, \gamma(s)} \,\ds 
     =
     \int_0^L \bar{m}(s)\,\ds + C_0 + C_L \,,
\end{array}
\end{equation}
which represents the overall rotational equilibrium of the rod, which need  not be added to the translational equilibrium \cref{eq:ContRodEquilFFExplicit}.

\subsection{Lagrangian of the equivalent rod}

The  equilibrium equations \eqref{eq:ContElemEquilExplicit} and the constitutive relations \eqref{eq:ContConstRelExplicit} 
for the homogeneous rod can be equivalently derived by considering a Lagrangian $\mathcal{L}$
\begin{equation}\label{lagrangianissimo}
    \mathcal{L}=\mathcal{E}-\mathcal{W}+\mathcal{M} \,,
\end{equation}
where $\mathcal{E}$ is the elastic energy of the equivalent rod, quadratic in the curvature $\theta'(s)$  and in the shear angle $\gamma(s)$,
\begin{equation}
    \mathcal{E} \roundc{\theta'(s), \gamma(s)} = \dfrac{1}{2}\int_{0}^L\squarea{EI \roundc{\theta'(s)}^2 
        + GA\ped{s} \roundc{\gamma(s)}^2} \ds \,,
\end{equation}
$\mathcal{W}$ is the work on the equivalent rod of the external distributed loads ($\bar{\Bq}(s)$ and $\bar{m}(s)$) and concentrated end loads ($\BF_0$, $C_0$, $\BF_L$,  and $C_L$)
\begin{equation}
    \mathcal{W} \roundc{\theta(s),\mathbf{u}(s)} = \int_{0}^L \squarec{\bar{\Bq}(s)\cdot \Bu(s)+\bar{m}(s)\theta(s)} \ds
    + \BF_0 \cdot \Bu_0 + C_0\,\theta_0
    + \BF_L\cdot\Bu_L + C_L\,\theta_L \,,
\end{equation}
and $\mathcal{M}$ is the contribution of the Lagrange multiplier $\mathbf{N}(s)$ (to be mechanically interpreted as the force internal to the equivalent rod),  needed to enforce the kinematic constraint between $\mathbf{u}(s)$ and the primary kinematic fields $\theta(s)$ and $\gamma(s)$,
\begin{equation}
    \mathcal{M} \roundc{\theta(s),\gamma(s),\mathbf{u}(s)} = \int_{0}^L \BN(s) \cdot \squarec{\Bu'(s)-\roundc{\roundb{1-\alpha} \Ba_1(s) + \alpha\, \Bb_1(s)-\Be_1}} \ds \,,
\end{equation}
where $\Ba_1(s)$ and $\Bb_1(s)$ are the unit vectors rotated at the angles  $\theta(s)$ and $\theta(s) + \gamma(s)/\alpha$ with respect to $\mathbf{e}_1$, Eqs.~\eqref{eq:ContSectVectors}.

After integration by parts, the first variation $\delta\mathcal{L}$ of the Lagrangian $\mathcal{L}$ \eqref{lagrangianissimo} is given by
\begin{equation}
    \begin{array}{ll}
    \delta\mathcal{L} =&\displaystyle
        -\int_{0}^L\squarec{EI \theta''(s)+\mathbf{N}(s)\cdot \roundc{\roundb{1-\alpha} \Ba_2(s) + \alpha\, \Bb_2(s)} + \bar{m}(s)}\delta\theta(s) \,\ds \\[2ex]
    &\displaystyle+ \int_{0}^L\squarec{GA\ped{s} \gamma(s)-
        \BN(s)\cdot \Bb_2(s)}\delta\gamma(s) \,\ds
    - \int_{0}^L\squarec{\BN'(s)+\bar{\Bq}(s)} \cdot \delta\Bu(s) \,\ds \\[2ex]
    &\displaystyle
    +\squarec{\BN(L) - \BF_L} \cdot \delta\Bu_L
    - \squarec{\BN(0) + \BF_0} \cdot \delta\Bu_0
    + \squarec{EI\, \theta'(L)-C_L}\delta\theta_L
    - \squarec{EI\, \theta'(0)+C_0}\delta\theta_0 \,,
    \end{array}
\end{equation}
and its vanishing for every admissible variation $\delta\theta(s)$, $\delta\gamma(s)$, and $\delta\Bu(s)$ implies the equilibrium equations \eqref{eq:ContElemEquilExplicit}, the constitutive relations \eqref{eq:ContConstRelExplicit},  and the boundary conditions \eqref{eq:ContBoundEquil} and \eqref{eq:ContRodEquilBC}.

\subsection{The Engesser rod as the special case \texorpdfstring{$\alpha\to 1$}{alpha -> 1}}\label{sec:EngesserGeneral}

The micromechanical parameter $\alpha$, implementing into the equivalent rod  the ratio between the extension of the four-bar linkage and the length of the unit cell, Fig.~\ref{fig:DiscreteLinkageKinem}a, plays a key role in defining the direction of the internal shear force in the equivalent continuous rod.

In the limit case of $\alpha \to 1$, the continuous rod specializes to the Engesser rod model. Indeed, in such a  limit the rod kinematics is characterized by an inextensible rod axis, \cref{eq:ContStretch},  an axis deviation angle $\chi(s)$ coincident with the shear angle $\gamma(s)$, Eqs.~\eqref{eq:ContAxisDeviationAngle}, and therefore a rotation angle of the tangent to the rod axis $\varphi(s)$ linear in the shear angle $\gamma(s)$,  \cref{eq:ContAxisRotationAngle}, namely
\begin{equation}\label{prelasteq}
    \lambda(s) = 1,\qquad 
    \chi(s)=\gamma(s),\qquad
    \varphi(s) = \theta(s) + \gamma(s).
\end{equation} 
Consequently, the unit vectors $\Bb_1(s)$ and $\Bb_2(s)$, defining the normal and the shear components of the internal force  $\BN(s)$, coincide with the tangent $\Bt(s)$ and the normal $\Bn(s)$ to the rod axis,  Eqs.~\eqref{eq:ContAxisVectors}, and simplify as
\begin{equation}\label{eq:EngesserAxisVectors}
    \Bt(s)  = \Bb_1(s)= \cos{\varphi(s)} \, \Be_1 + \sin{\varphi(s)} \, \Be_2 \,,
    \qquad
    \Bn(s) = \Bb_2(s)=  -\sin{\varphi(s)} \, \Be_1 + \cos{\varphi(s)} \, \Be_2 \,.
\end{equation}

Moreover, \cref{eq:ContRodEquilDAE}\ped{2} can be solved for the shear angle $\gamma(s)$, leading to
\begin{equation}\label{eq:EngesserShearAngle}
    \gamma(s) = -\dfrac{\roundc{\BQ(s) + \BF_L} \cdot 
    \roundc{\sin{\varphi(s)}\,\Be_1 - \cos{\varphi(s)}\,\Be_2}}{GA\ped{s}} \,,
\end{equation}
so that the DAE \eqref{eq:ContRodEquilDAE} can be reduced to a second-order differential equation, nonlinear in the unknown function $\varphi(s)$, 
\begin{equation}\label{eq:EngesserSingleEq}
\begin{array}{l}
    EI \squaref{\varphi(s) + \dfrac{\roundc{\BQ(s) + \BF_L} \cdot 
    \roundc{\sin{\varphi(s)}\,\Be_1 - \cos{\varphi(s)}\,\Be_2}}{GA\ped{s}}}''
    \\[3ex]
    \hspace{35ex}
    - \roundc{\BQ(s) + \BF_L} \cdot \roundc{\sin{\varphi(s)}\,\Be_1 - \cos{\varphi(s)}\,\Be_2} + \bar{m}(s) =0 \,,
\end{array}
\end{equation}
to be complemented by the boundary conditions at the rod ends, Eqs.~\eqref{eq:ContRodEquilBC}, 
\begin{equation}\label{eq:EngesserBoundaryC0}
    \squaref{\varphi(s) + \dfrac{\roundc{\BQ(s) + \BF_L} \cdot 
    \roundc{\sin{\varphi(s)}\,\Be_1 - \cos{\varphi(s)}\,\Be_2}}{GA\ped{s}}}'_{s = 0} =- \frac{C_{0}}{EI} \,,
\end{equation}
and
\begin{equation}\label{eq:EngesserBoundaryCL}
    \squaref{\varphi(s) + \dfrac{\roundc{\BQ(s) + \BF_L} \cdot 
    \roundc{\sin{\varphi(s)}\,\Be_1 - \cos{\varphi(s)}\,\Be_2}}{GA\ped{s}}}'_{s = L} = \frac{C_L}{EI} \,.
\end{equation}

The function $\varphi(s)$, solution of the differential equation \eqref{eq:EngesserSingleEq}, can be used in \cref{eq:EngesserShearAngle} to obtain the shear angle $\gamma(s)$, and the cross-section rotation angle can be evaluated as $\theta(s) = \varphi(s) - \gamma(s)$. The derivative $\varphi'(s)$ represents the rod axis curvature, and the derivative $\theta'(s)$ can be used in \cref{eq:ContConstRelExplicit}\ped{2} to obtain the bending moment $M(s)$. Moreover, the internal shear force, directed along $\Bn(s) = \Bb_2(s)$, is obtained from $\gamma(s)$ using \cref{eq:ContInternalShearForce}, and the normal component of the internal force, in the direction of $\Bt(s) = \Bb_1(s)$, \cref{eq:EngesserAxisVectors}\ped{1}, can be obtained through a specialization of  \cref{eq:ContInternalNormalForce} to
\begin{equation}
    \BN(s) \cdot \Bt(s) = - \roundc{\cos{\varphi(s)}\, \Be_1 + \sin{\varphi(s)}\, \Be_2} \cdot \squaree{\BF_0 + \int_0^s \bar{\Bq}(\sigma) \,\dsigma} \,.
\end{equation}

Finally, the horizontal and vertical components of the displacement field can be obtained through an integration of  Eqs.~\eqref{eq:ContDispDE}, which simplify as
\begin{equation}\label{eq:EngesserDispDE}
    u_1'(s) = \cos{\varphi(s)} - 1 \,,
    \qquad
    u_2'(s) = \sin{\varphi(s)} \,.
\end{equation}

It should be noted that the above-derived Eqs.\,(\ref{eq:EngesserSingleEq})--(\ref{eq:EngesserDispDE}) govern the equilibrium  at large deformation  of the Engesser rod subject to  general loading conditions, never introduced before.  

\paragraph{Recovering the Euler rod.}

The model of an inextensible and unshearable rod, namely, the Euler rod, can be obtained as a special case of the Engesser rod for $GA\ped{s} \to \infty$, obtained in the limit of $k_\beta\to\infty$, \cref{eq:ContSectionStiffness}\ped{1}. This condition  implies a null shear angle,  \cref{eq:EngesserShearAngle}, and in turn a rotation angle $\varphi(s)$ of the tangent to the rod axis coincident with the rotation angle of the cross-section $\theta(s)$, 
\begin{equation}\label{lasteqn}
    \gamma(s) = 0 \,, \qquad \varphi(s)=\theta(s) \,.
\end{equation}
It follows that the unit vectors tangent $\Bt(s)$ and normal $\Bn(s)$ to the rod axis become coincident with the unit vectors orthogonal $\Ba_1(s)$ and parallel $\Ba_2(s)$ to the cross-section, 
\begin{equation}\label{eq:EulerAxisVectors}
    \Bt(s) =\Ba_1(s) = \cos{\theta(s)} \, \Be_1 + \sin{\theta(s)} \, \Be_2 \,,
    \qquad
    \Bn(s) =\Ba_2(s)= -\sin{\theta(s)} \, \Be_1 + \cos{\theta(s)} \, \Be_2 \,.
\end{equation}

The differential equation \eqref{eq:EngesserSingleEq} further specializes to the well-known  Euler elastica equation \cite{CISM2019} 
\begin{equation}\label{eq:EulerSingleEq}
\begin{array}{l}
    EI\, \theta''(s)
    - \roundc{\BQ(s) + \BF_L} \cdot \roundc{\sin{\theta(s)}\,\Be_1 - \cos{\theta(s)}\,\Be_2} + \bar{m}(s) =0 \,,
\end{array}
\end{equation}
to be complemented with the boundary conditions $EI\, \theta'(0) =- C_0$ and $EI\, \theta'(L) = C_L$. The differential equations \eqref{eq:EngesserDispDE} defining the components of the displacement become
\begin{equation}\label{eq:EulerDispDE}
    u_1'(s) = \cos{\theta(s)} - 1 \,,
    \qquad
    u_2'(s) = \sin{\theta(s)} \,.
\end{equation}

It is observed that the Euler rod model can also be obtained from our model of equivalent continuous rod  in the limit case of $\alpha \to 0$, which implies an infinite shear stiffness, \cref{eq:ContSectionStiffness}\ped{1}, $GA\ped{s}\to\infty$ and  the shear angle $\gamma(s)$ vanishes, \cref{eq:ContRotationsLimit}\ped{2}. Moreover, despite the ratio $\gamma(s)/\alpha$ becomes undefined, Eqs.~\eqref{eq:ContStretch} and \eqref{eq:ContAxisDeviationAngle} provide $\lambda(s) = 1$ and $\chi(s) = 0$, so that the rod becomes both unstretchable and unshearable, \cref{prelasteq}\ped{1} and \cref{lasteqn}, and the differential equations \eqref{eq:ContDispDE} reduce to Eqs.~\eqref{eq:EulerDispDE}.

Finally, the differential equation \cref{eq:ContRodEquilDAE}\ped{1} becomes \cref{eq:EulerSingleEq}, while the algebraic equation (\ref{eq:ContRodEquilDAE})\ped{2}, representing the equilibrium of the shear internal force, does not longer apply when $\alpha \to 0$. 

\section{Bifurcation of the discrete and the equivalent continuous models for cantilever and simply supported schemes}

\begin{figure}[h]
    \centering
    \includegraphics[width=170mm]{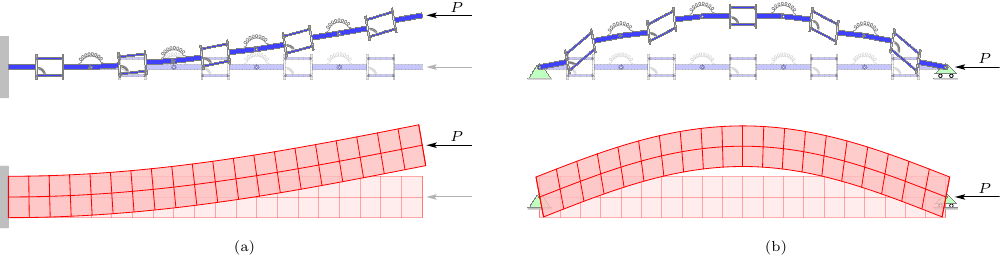}
    \caption{The cantilever (a) and simply supported (b)  structural schemes used in the bifurcation analysis of the discrete (upper part) and equivalent continuous models of the rod (lower part). 
    The first buckling mode for a discrete system (with $n=5$) and for its equivalent continuous counterpart is superimposed to the undeformed configuration. Results are for $\alpha=0.3$, and $\zeta=20$. 
    }
    \label{fig:BucklingSchemes}
\end{figure}

The bifurcation conditions for the straight configuration are analytically solved for both the discrete chain model and its equivalent continuous counterpart, subject to an axial load $P$. 
Two different boundary conditions are considered, corresponding to cantilever and simply supported schemes, Fig.~\ref{fig:BucklingSchemes} (henceforth, the negative load $P$ is drawn on the structure by making explicit its true direction). 
The obtained bifurcation loads and modes for the discrete chains in the limit of an infinite number of elements ($n\to\infty$) are shown  to be coincident with those of the corresponding equivalent rod, confirming the effectiveness of the homogenized model. The analysis is performed with reference to the dimensionless load $p$ and the (shear vs bending) stiffness ratio $\zeta$
\begin{equation}\label{eq:ForceStiffnessNormalization}
    p = \dfrac{P L^2}{EI} = \dfrac{P a n^2}{K} \,,
    \qquad
    \zeta = \dfrac{GA\ped{s} L^2}{EI} = \dfrac{k_\beta n^2}{\alpha^2 K}.
\end{equation}

The bifurcation conditions for the two different boundary value problems  evidence  strict analogies, as follows.
\begin{itemize}
    \item The bifurcation loads for the discrete chain (for the continuous equivalent model) occur for two \lq twin' sequences $p_m^+$ and $p_m^-$ of $n$ (of infinite) values given by pairs of bifurcation conditions associated to the same mode number $m$, but belonging to the different sequences. The critical $m$--th mode of each of the twin sequences share the same critical shape for the rod axis, but obtained through a different critical shear deformation contribution. 
    
    \item The values sets $p_m^+$ and $p_m^-$ of the \lq twin' sequences satisfy a strict inequality, so all the values of the first sequence ($p_m^+$) are smaller than all the corresponding values of the second sequence ($p_m^-$). The two sequences are separated from each other by a \lq transition' bifurcation load $p_0$, which is independent of the number $n$ of  elements characterizing the chain and is coincident with the \lq transition load' of the equivalent rod, 
\begin{equation}
\label{pizzone}
...<
p^-_{m+1}<
p^-_{m}
<...
<p^-_2
< p^-_1
<p_0\leq p^*
<...<
p^+_{m+1}
<
p^+_{m}
<...
<p^+_2
<p^+_1<0 \,. 
\end{equation}
Note that an additional characteristic load, namely, $p^*$,  appears in \cref{pizzone}. This is a value for which the equations degenerate  and will be later demonstrated to represent an accumulation value for the bifurcations of the equivalent rod. The value $p^*$ corresponds to the buckling load $P^*$ of the four-bar linkage,
\begin{equation}
\label{eq:ContAccumulationLoad}
    p^*=-\alpha\zeta
    \qquad\Longleftrightarrow
    \qquad
    P^*=-\dfrac{k_\beta}{\alpha\, a} \,.
\end{equation}

\item In the limit of infinite shear to bending stiffness ratio, the first sequence approaches that characterizing Euler buckling of the inextensible and unshearable elastica, while both the transition load $p_0$ and the second sequence ($p_m^-$) diverge to infinity, therefore becoming mechanically irrelevant. 

\end{itemize}

The transition  load $p_0$ for the two boundary value problems is associated with  two substantially different critical modes, 
\begin{itemize}
\item a \lq bookshelf-like' shape for the simply supported scheme, 
where the rod axis remains straight;
\item a transverse displacement jump occurring at the clamp in an otherwise straight rod axis for the cantilever scheme.
\end{itemize} 

\subsection{Discrete chain models}
\label{discretino}

A cantilever chain and a simply supported chain of discrete elements, as illustrated in Fig.~\ref{fig:DiscreteLinkageKinem}a, are shown in Fig.~\ref{fig:BucklingSchemes} and are loaded at the $n$--th node through a  force $P$ parallel to  $\Be_1$ (therefore positive when  tensile in the straight configuration). The distributed loads and the inner nodal forces are null, so that the resultant of external forces $\BQ^i$ vanishes, \cref{eq:DiscGeneralForceLoadExplicit}. In both the considered structural schemes, the vertical force applied on the $n$--th node is null. Therefore, from Eq.~\eqref{eq:DiscRodEquilFFExplicit}, the forces applied on both ends of the rods are 
\begin{equation}\label{eq:DiscrExtremityForces}
        \BF_n =-\BF_0 =  P \Be_1 \,,
\end{equation}
and Eqs.~\eqref{eq:DiscRodEquilExplicit} simplify to
\begin{equation}\label{eq:DiscApplicationExactEq}
\left\{
\begin{array}{l}
    K \roundb{\theta^1 - \theta^{2}} + P a \squarec{\roundb{1-\alpha} \sin{\theta^1} +  \alpha \sin\roundb{\theta^1 + \beta^1}} - C_0 = 0 \,, \\[2ex]
    K \roundb{-\theta^{i-1} + 2 \theta^i - \theta^{i+1}} + P a \squarec{\roundb{1-\alpha} \sin{\theta^i} +  \alpha \sin\roundb{\theta^i + \beta^i}}  = 0 \,,
    \quad i = 2, \ldots, n-1 \,, \\[2ex]
    K \roundb{\theta^{n} - \theta^{n-1}} + P a \squarec{\roundb{1-\alpha} \sin{\theta^n} + \alpha \sin\roundb{\theta^n + \beta^n}} = 0 \,, \\[2ex]
    k_\beta \beta^i + P a \alpha  \sin\roundb{\theta^i + \beta^i} = 0 \,,
    \quad i = 1, \ldots, n \,,
\end{array}
\right.
\end{equation}
which have to be further specialized to keep into account the boundary conditions, different in the two schemes.
For the analysis of these structures, 
it is instrumental to  collect the configuration parameters $\theta^i$ and $\beta^i$, with $i=2,\ldots,n$, into the $\roundb{n-1}$-dimensional arrays
\begin{equation}\label{eq:DiscConfigParamArray}
     \Btheta = \squareb{\theta^2 \, \ldots \, \theta^n}\tra \,,
     \qquad
     \Bbeta = \squareb{\beta^2 \ldots \beta^n}\tra \,.
\end{equation}

\subsubsection{Cantilever chain}

For a cantilever chain, Fig.~\ref{fig:BucklingSchemes}a (upper part), the moment applied at the $n$--th node is null, $C_n=0$, and the bending moment $C_0$ at the clamp can be evaluated using the global rotational equilibrium, \cref{eq:DiscRodEquilMMExplicit}, together with \cref{eq:DiscrExtremityForces} and the boundary condition $\theta^1 = 0$, thus obtaining
\begin{equation}\label{eq:CantilevMoment}
    C_0 = P a \squared{\alpha \sin\beta^1 + \sum_{i=2}^n \squarec{\roundb{1-\alpha} \sin{\theta^i} + \alpha \sin\roundb{\theta^i + \beta^i}}} \,.
\end{equation}

In order to analyse the bifurcation of the structure from its undeformed state ($\theta^i = \beta ^ i = 0$), the linear approximations   
$\sin{\theta^i} \approx \theta^i$ and 
$\sin\roundb{\theta^i+\beta^i} \approx \theta^i+\beta^i$ are considered. Furthermore, noting that \cref{eq:CantilevMoment} is the sum of Eqs.~\eqref{eq:DiscApplicationExactEq}\ped{1}--\eqref{eq:DiscApplicationExactEq}\ped{3}, the first equation of \eqref{eq:DiscApplicationExactEq} becomes redundant, so that the cantilever chain may be analysed by considering only equations \eqref{eq:DiscApplicationExactEq}\ped{2}--\eqref{eq:DiscApplicationExactEq}\ped{4}, specifically assembled in the matrix form
\begin{equation}\label{eq:DiscCantilevLinearMat}
    \BA \Btheta + \dfrac{p}{n^2} \roundb{\Btheta + \alpha \, \Bbeta} = \Bnull \,,
    \quad
     p \Btheta + \roundb{\alpha\zeta + p} \Bbeta = \Bnull \,,
     \quad
    \roundb{\alpha\zeta + p} \beta^1 = 0 \,,
\end{equation}
where $\BA$ is the tridiagonal and symmetric matrix of dimension $\roundb{n-1}$ defined as 
\begin{equation}\label{eq:DiscNormalizedElasticMat}
    \BA = 2 \BI - \BF_{n-1} \,, 
\end{equation}
where 
\begin{equation}\label{eq:EigenMatrixFF}
\setlength\arraycolsep{1.2pt}
\BF_k=
\left[
\phantom{\hspace{-1.4em}\begin{matrix} \ddots\\[-.6em] \ddots\\[-.6em] \ddots\\[-.6em] \ddots\\[-.6em] \ddots\\[-.6em] 0 \end{matrix}}
\right.
    \underbrace{
    \begin{matrix}
        0 & 1 & 0 & \cdots & \cdots & 0 \\[-.6em]
        1 & 0 & 1 & \ddots & & \vdots \\[-.6em]
        0 & 1 & \ddots  & \ddots & \ddots & \vdots \\[-.6em]
        \vdots & \ddots & \ddots & \ddots & 1 & 0 \\[-.6em]
        \vdots & & \ddots & 1 & 0 & 1 \\
        0 & \cdots & \cdots & 0 & 1 & 1
    \end{matrix}
    }_{k}
\left.
\phantom{\hspace{-1.4em}\begin{matrix} \ddots\\[-.6em] \ddots\\[-.6em] \ddots\\[-.6em] \ddots\\[-.6em] \ddots\\[-.6em] 0 \end{matrix}}
\right]
\left.
\phantom{\hspace{-1.4em}\begin{matrix} \ddots\\[-.6em] \ddots\\[-.6em] \ddots\\[-.6em] \ddots\\[-.6em] \ddots\\[-.6em] 0 \end{matrix}}
\right\}{\scriptstyle{k}}
\,.
\end{equation}

The  solution $\Bbeta$ and $\Btheta $ to  Eqs.~\eqref{eq:DiscCantilevLinearMat} can be found by confining attention to the case $p \neq p^*$.
More specifically, when $p \neq p^*$, \cref{eq:DiscCantilevLinearMat}\ped{3} provides $\beta^1 = 0$ and \cref{eq:DiscCantilevLinearMat}\ped{2} can be inverted to give
\begin{equation}\label{eq:DiscCantilevBeta2Theta}
    \Bbeta = -\dfrac{p}{\alpha\zeta + p} \Btheta \,,
\end{equation}
an equation that can be used in \cref{eq:DiscCantilevLinearMat}\ped{1} to assemble the following eigenvalue problem
\begin{equation}\label{eq:DiscCantilevEigenSystem}
    \rounde{\BA - \dfrac{\omega^2}{n^2} \BI}\Btheta = \Bnull \,,
\end{equation}
where $\omega^2$ is defined as
\begin{equation}\label{eq:DiscCanilevQuadEigen}
    \omega^2 = -p\,\dfrac{\alpha\zeta + \roundb{1-\alpha}p}{\alpha\zeta + p}  \,.
\end{equation}

By virtue of \cref{eq:DiscNormalizedElasticMat}, the eigenvalues of the matrix $\BA$ can be expressed in the form $2 - \lambda_m^{(n-1)}$, being $\lambda_m^{(n-1)}$ the $m$--th eigenvalue of $\BF_{n-1}$. In particular, the eigenvalues of the matrix $\BF_k$ for any $k \ge 1$ are (see Appendix \ref{sec:MatrixEigenFF} for details)
\begin{equation}\label{eq:EigenValuesFF}
    \lambda_m^{(k)} = 2 \cos\rounde{\dfrac{\roundb{2 m - 1}\pi}{2 k +1}} \,,
    \qquad m = 1, \ldots, k \,,
\end{equation}
whence, setting $k = n-1$, the solution of the eigenvalue problem,  \cref{eq:DiscCantilevEigenSystem}, becomes 
\begin{equation}\label{eq:DiscCantilevEulerCrit}
    \omega^2_m(n) = 4 n^2 \sin^2\left(\dfrac{\pi\roundb{2 m - 1}}{2\roundb{2 n - 1}}\right)\,, \qquad m = 1, \ldots, n-1 \,.
\end{equation}
By inverting \cref{eq:DiscCanilevQuadEigen} and assuming $\omega^2 = \omega^2_m(n)$, the \emph{twin} sequences of (negative) bifurcation loads $p_m^+(n)$ and $p_m^-(n)$ are obtained  
\begin{equation}\label{eq:DiscCantilevCritLoads}
\left.
\begin{array}{l}
p_m^+(n) \\ [1ex]
p_m^-(n)
\end{array} 
\right\}
    = \dfrac{-\roundc{\alpha\zeta + \omega^2_m(n)} \pm \sqrt{\roundc{\alpha\zeta + \omega^2_m(n)}^2 - 4\alpha\roundb{1-\alpha} \zeta \omega^2_m(n)}}{2\roundb{1-\alpha}} \,,
    \qquad
    m = 1, \ldots, n-1\,.
\end{equation}

In addition to the twin sequences of bifurcation loads $p^+_m$ and $p^-_m$, \cref{eq:DiscCantilevCritLoads}, an additional  bifurcation condition exists, associated with the negative  \emph{transition load} $p_0$, which can be obtained from Eqs.~\eqref{eq:DiscCantilevLinearMat} 
as
\begin{equation}\label{eq:DiscCantileverLocalShape}
    p_0=-\alpha\zeta \,, \quad
    \beta^1 = c  \,,
    \quad
    \theta^i =    \beta^i = 0 \,,
    \qquad
    i=2,...,n \,,
\end{equation}
where $c$ is an arbitrary amplitude, defining a jump in the axis displacement (see the inset labelled \lq Transition mode' in Fig.~\ref{fig:cantilever}).

The transition load $p_0$ coincides\footnote{This coincidence does not occur for the simply supported chain, as will be shown later.} with $p^*$, \cref{eq:ContAccumulationLoad}, namely,
\begin{equation}\label{eq:DiscCantilevLocalLoad}
    p_0 = p^* \,,
\end{equation}
and corresponds to the vanishing of the coefficient multiplying $\beta^1$ in \cref{eq:DiscCantilevLinearMat}\ped{3}. 
Note that $p_0$ is intermediate between the twin critical loads sequences, according to inequality \eqref{pizzone}, as
$$
p^-_{n-1}
<...
<p^-_2
< p^-_1
<p_0
<p^+_{n-1}
<...
<p^+_2
<p^+_1<0 \,.
$$

\subsubsection{Simply supported chain}\label{sec:DiscSimplysupBuckling}

For a simply supported chain, Fig.~\ref{fig:BucklingSchemes}b (upper part),  the bending moments at both ends of the rod vanish, $C_0 = C_n = 0$, and taking into account \cref{eq:DiscrExtremityForces}, the global rotational equilibrium, \cref{eq:DiscRodEquilMMExplicit}, provides
\begin{equation}\label{eq:DiscSimplySuppKinematicBoundary}
    a \sum_{i=1}^n \squarec{\roundb{1-\alpha} \sin{\theta^i} + \alpha \sin\roundb{\theta^i + \beta^i}}= 0 \,,
\end{equation}
which is used in place of Eq.~\eqref{eq:DiscApplicationExactEq}\ped{1} in analysing the discrete chain. Note that \cref{eq:DiscSimplySuppKinematicBoundary} corresponds to the kinematic boundary condition $\Bu_n \cdot \Be_2  = 0$, where the displacement $\Bu_n \cdot \Be_2$ can be evaluated by specializing \cref{eq:DiscRodDisplacementExplicit}\ped{2} to $i = n$ and using the condition $\Bu_0 = \Bnull$.

The bifurcation of the chain is analysed by considering  once more the linearized approximations 
$\sin{\theta^i} \approx \theta^i$ and 
$\sin\roundb{\theta^i+\beta^i} \approx \theta^i+\beta^i$, so that with reference to the dimensionless parameters,  \cref{eq:ForceStiffnessNormalization}, and vectors $\Btheta$ and $\Bbeta$ defined by \cref{eq:DiscConfigParamArray}, equations \eqref{eq:DiscApplicationExactEq}\ped{2}--\eqref{eq:DiscApplicationExactEq}\ped{4} and \eqref{eq:DiscSimplySuppKinematicBoundary} can be assembled in the following matrix form
\begin{equation}\label{eq:DiscSimplySuppLinearMat}
\begin{array}{ll}
     -\theta^{1} \Be + \BA \Btheta + \dfrac{p}{n^2} \roundb{\Btheta + \alpha \, \Bbeta} = \Bnull \,,
    & \roundb{\theta^1 + \alpha \, \beta^1} + \Bunitary \cdot \roundb{\Btheta + \alpha \, \Bbeta} = 0 \,,\\ [3ex]
    p \theta^1 + \roundb{\alpha\zeta + p} \beta^1 = 0 \,,
    &  p \Btheta + \roundb{\alpha\zeta + p} \Bbeta = \Bnull \,,
    \end{array}
\end{equation}
where $\Be= \squareb{1 \enspace 0 \ldots 0}\tra$, $\Bunitary = \squareb{1 \enspace \ldots \enspace 1}\tra$, and $\BA$ is the tridiagonal symmetric matrix expressed by \cref{eq:DiscNormalizedElasticMat}.

Assuming $p \neq p^*$, \cref{eq:DiscSimplySuppLinearMat}$_4$, can be solved for $\Bbeta$ to again obtain \cref{eq:DiscCantilevBeta2Theta}, moreover \cref{eq:DiscSimplySuppLinearMat}\ped{3}, provides $\beta^1$, so that \cref{eq:DiscSimplySuppLinearMat}\ped{2} yields
\begin{equation}\label{eq:DiscSimplySuppConstrCond}
    \dfrac{\alpha\zeta + \roundb{1-\alpha}p}{\alpha\zeta + p} \roundb{\theta^1 + \Bunitary \cdot \Btheta} = 0 \,,
\end{equation}
an equation that is satisfied when either $\roundb{\theta^1 + \Bunitary \cdot \Btheta}$ or its coefficient vanish. In the former case, the relation $\theta^1 = -\Bunitary \cdot \Btheta$ can be used in \cref{eq:DiscSimplySuppLinearMat}\ped{1} along with \cref{eq:DiscCantilevBeta2Theta} to obtain the eigenvalue problem
\begin{equation}\label{eq:DiscSimplySuppEigenSystem}
    \rounde{\BB - \dfrac{\omega^2}{n^2} \BI}\Btheta = \Bnull \,,
\end{equation}
where $\BB = \BA + \Be \otimes \Bunitary$ and $\omega^2$ is the parameter defined by \cref{eq:DiscCanilevQuadEigen}.

The eigenvalue problem \cref{eq:DiscSimplySuppEigenSystem} is formally analogous to \cref{eq:DiscCantilevEigenSystem} governing the bifurcation of the cantilever chain, so that the same strategy is followed to solve it. 
In this case, the auxiliary matrix to be considered is
\begin{equation}
        \BH_k = \BF_{k} - \Be \otimes \Bunitary \,,
\end{equation}
where $\BF_{k}$ is still defined by Eq.~\eqref{eq:EigenMatrixFF} and therefore $\BH_k$  follows as
\begin{equation}\label{eq:EigenMatrixHH}
\setlength\arraycolsep{1.2pt}
\BH_k=
\left[
\phantom{\hspace{-1.4em}\begin{matrix} 0\\[-.2em] 0\\[-.6em] \ddots\\[-.6em] \ddots\\[-.6em] \ddots\\[-.6em] \ddots\\[-.1em] 0 \end{matrix}}
\right.
    \underbrace{
    \begin{matrix}
        -1 & 0 & -1 & \cdots &\cdots & \cdots & -1 \\[-.2em]
        1 & 0 & 1 & 0 & \cdots & \cdots& 0 \\[-.6em]
        0 & 1 & 0 & \ddots & \ddots & & \vdots \\[-.6em]
        \vdots & \ddots & \ddots& \ddots  & \ddots & \ddots & \vdots \\[-.6em]
        \vdots &  & \ddots & \ddots & \ddots & 1 & 0 \\[-.6em]
        \vdots & & &\ddots & 1 & 0 & 1 \\[-.1em]
        0 & \cdots & \cdots & \cdots & 0 & 1 & 1
    \end{matrix}
    }_{k}
\left.
\phantom{\hspace{-1.4em}\begin{matrix} 0\\[-.2em] 0\\[-.6em] \ddots\\[-.6em] \ddots\\[-.6em] \ddots\\[-.6em] \ddots\\[-.1em] 0 \end{matrix}}
\right]
\left.
\phantom{\hspace{-1.4em}\begin{matrix} 0\\[-.2em] 0\\[-.6em] \ddots\\[-.6em] \ddots\\[-.6em] \ddots\\[-.6em] \ddots\\[-.1em] 0 \end{matrix}}
\right\}{\scriptstyle{k}}
\,,
\end{equation}
with eigenvalues expressed as (see Appendix \ref{sec:MatrixEigenHH} for details)
\begin{equation}\label{eq:EigenValuesHH}
    \lambda_m^{(k)} = 2 \cos\rounde{\dfrac{m \pi}{k+1}} \,, \qquad m = 1, \ldots, k \,.
\end{equation}
Specializing \cref{eq:EigenMatrixHH} to $k=n-1$, one can easily recognize that $\BB = 2 \BI - \BH_{n-1}$, whence the $m$--th eigenvalue of $\BB$ is $\omega_m^2(n)/n^2 = 2 - \lambda_m^{(n-1)}$. The latter eigenvalue, keeping into account \cref{eq:EigenValuesHH}, can be reduced to
\begin{equation}\label{eq:DiscSimplySuppEulerCrit}
   \omega_m^2(n) = 4 n^2 \sin^2\roundd{\dfrac{m \pi}{2 n}} \,, \qquad m = 1, \ldots, n-1 \,.
\end{equation}

Furthermore, the definition of $\omega^2$ in \cref{eq:DiscCanilevQuadEigen} applies to both the cantilever and the simply supported chain, so that the bifurcation loads can be evaluated again using \cref{eq:DiscCantilevCritLoads}, but referring to \cref{eq:DiscSimplySuppEulerCrit} for the values of $\omega_m^2(n)$ specific to the simply supported chain.

Similarly to the cantilever chain, the bifurcation of the simply supported chain is  characterized by a \emph{twin} sequence of loads, $p^+_m$ and $p^-_m$, separated by an additional bifurcation condition occurring for the transition load $p_0$. This condition arises when the constraint expressed by \cref{eq:DiscSimplySuppConstrCond} is satisfied because the coefficient $\alpha\zeta + \roundb{1-\alpha}p$  vanishes, whence
\begin{equation}\label{eq:DiscSimplySuppLocalLoad}
    p_{0} = -\dfrac{\alpha\zeta}{1-\alpha} \,,
\end{equation}
while \cref{eq:DiscSimplySuppLinearMat}\ped{1} becomes $\BA \Btheta = \theta^1 \Be$, whose solution provides the bifurcation mode in the form
\begin{equation}\label{eq:DiscSimplySuppLocalShape}
    \theta^i_{0} = c \,,
    \quad
    \beta^i_{0} = -\dfrac{c}{\alpha}  \,,
    \qquad
    i = 1, \ldots, n \,,
\end{equation}
where $c$ is an arbitrary amplitude. The bifurcation mode, \cref{eq:DiscSimplySuppLocalShape}, defines a \lq bookshelf-like' configuration (see the inset labelled \lq Transition mode' in Fig.~\ref{fig:SimplySuppBifurcation}).

\subsection{The continuous rod 
equivalent to the chain 
and the match of the two models on the bifurcation loads and modes when \texorpdfstring{$n\to\infty$}{n -> infinity}}
\label{equivalentino}

The homogenized rod model of the chains 
is analysed in the configuration shown in 
Fig.~\ref{fig:BucklingSchemes}. In particular, the  force $P$ along $\Be_1$ is applied at $s=L$, while the distributed load $\bar{\Bq}(s)$ and couple $\bar{m}(s)$ are null, implying the vanishing of the resultant external force $\BQ(s)$, \cref{eq:ContGeneralForceLoadMaterial}.

The forces at the ends of the rod are $\BF_L = P \Be_1$ and $\BF_0 = -P\Be_1$, satisfying Eq.~\eqref{eq:ContRodEquilFFExplicit}, and the differential-algebraic system of equations \eqref{eq:ContRodEquilDAE}  reduces to
\begin{equation}\label{eq:ContApplicationExactEq}
\left\{
\begin{array}{l}
    EI \, \theta''(s) -  P \,\roundb{1-\alpha} \sin{\theta(s)}  + \alpha\,GA\ped{s} \, \gamma(s) = 0 \,,
        \\[2ex]
    GA\ped{s} \, \gamma(s) + P \sin\roundc{\theta(s) + \gamma(s)/\alpha} = 0 \,,
\end{array}
\right.
\end{equation}
with the boundary conditions, Eqs.~\eqref{eq:ContRodEquilBC}, and the solvability condition \eqref{eq:ContRodEquilMMExplicit} which becomes
\begin{equation}\label{eq:ContApplicationExactMM}
     P \int_0^L \roundb{1-\alpha} \sin{\theta(s)} \,\ds 
     - \int_0^L \alpha\,GA\ped{s} \, \gamma(s) \,\ds =  EI\, \left[\theta'(L)  -  \theta'(0)\right] \,.
\end{equation}

\subsubsection{Cantilever configuration for the equivalent rod}

For the cantilever rod depicted in Fig.~\ref{fig:BucklingSchemes}a (lower part), the rotation at the clamp vanishes, while at the free end Eq.~\eqref{eq:ContRodEquilBC}\ped{2} must be used with the loading condition $C_L=0$, leading to the boundary conditions for the unknown rotation  field $\theta(s)$
\begin{equation}\label{eq:ContCantilevBoundaryCond}
    \theta(0) = 0 \,,
    \qquad
    \theta'(L) = 0 \,.
\end{equation}
which supplement the system of equations \cref{eq:ContApplicationExactEq}.  Eq.~\eqref{eq:ContRodEquilBC}\ped{1} can be used to evaluate the bending moment at the clamp when the function $\theta(s)$ and its derivative $\theta'(s)$ have been determined.

In order to use the dimensionless parameters in \cref{eq:ForceStiffnessNormalization} for the continuous rod, the dimensionless coordinate $\xi =s/L \in \squareb{0,1}$ is introduced. The bifurcation condition is obtained through a linearization of Eqs.~\eqref{eq:ContApplicationExactEq} in the form
\begin{equation}\label{eq:ContCantilevLinearEq}
    \theta''(\xi) -  p \,\roundb{1-\alpha} \,\theta(\xi)  + \alpha\zeta \, \gamma(\xi) = 0 \,, 
    \qquad
    \alpha p \, \theta(\xi) + \roundb{\alpha\zeta + p}  \gamma(\xi) = 0 \,,
\end{equation}
where the symbol $'$ denotes henceforth the derivative of the function with respect to the relevant variable.

Assuming $p \neq p^*$, \cref{eq:ContAccumulationLoad}, the algebraic equation \eqref{eq:ContCantilevLinearEq}\ped{2} can be solved for $\gamma(\xi)$ as
\begin{equation}\label{eq:ContCantilevGamma2Theta}
    \gamma(\xi) = -\dfrac{\alpha p}{\alpha\zeta + p} \theta(\xi) \,,
\end{equation}
which substituted in \cref{eq:ContCantilevLinearEq}\ped{1} provides
\begin{equation}\label{eq:ContCantilevDiffEq}
    \theta''(\xi)  + \omega^2 \, \theta(\xi)  = 0 \,,
\end{equation}
where the parameter $\omega^2$ is again given by \cref{eq:DiscCanilevQuadEigen}, introduced for the bifurcation analysis of the discrete chains.

Note that equation (\ref{eq:ContCantilevDiffEq}) is identical to the homogeneous differential problem ruling Euler's buckling for the inextensible and unshearable elastica except for the definition of $\omega$, \cref{eq:DiscCanilevQuadEigen}. Therefore, while the bifurcation loads result different, the bifurcation modes in terms of cross-section rotation $\theta(s)$ of the present equivalent shearable model  are identical with those corresponding to the  inextensible and unshearable elastica, namely
\begin{equation}\label{eq:ContCalilevGeneralSol}
    \theta(\xi) = A \sin\roundb{\omega \xi} + B \cos\roundb{\omega \xi} \,,
\end{equation}
where the boundary conditions, \cref{eq:ContCantilevBoundaryCond}, 
imply $B = 0$ and $\omega \cos{\omega} = 0$. It follows that, 
beside the trivial condition $\omega = 0$ (implying  $\theta(\xi) = \gamma(\xi)=0$), 
non-trivial solutions exist only when  $\cos{\omega} = 0$, namely, 
\begin{equation}\label{eq:ContCantilevEulerCrit}
    \omega_m^2 = \dfrac{\roundb{2m-1}^2\pi^2}{4} \,,
        \qquad m = 1, 2, \ldots \,,
\end{equation}
and, setting $\omega^2 = \omega^2_m$ in \cref{eq:DiscCanilevQuadEigen}, the bifurcation loads are obtained as \emph{twin} sequences of negative values $p_m^+$ and $p_m^-$ as
\begin{equation}\label{eq:ContCantilevCritLoads}
\left.
\begin{array}{l}
p_m^+ \\ [1ex]
p_m^-
\end{array} 
\right\}
    = \dfrac{-\roundc{\alpha\zeta + \omega^2_m} \pm \sqrt{\roundc{\alpha\zeta + \omega^2_m}^2 - 4\alpha\roundb{1-\alpha} \zeta \omega^2_m}}{2\roundb{1-\alpha}} \,,
    \qquad
    m = 1, 2, \ldots, \,.
\end{equation}
which are formally the same as \cref{eq:DiscCantilevCritLoads}, but now with the mode number $m$ unbounded from the above. Moreover,  Eqs.~(\ref{eq:ContCalilevGeneralSol}) and  (\ref{eq:ContCantilevGamma2Theta}), 
lead to the bifurcation modes 
\begin{equation}\label{eq:ContCantilevBifurcationShapes}
    \theta_m(\xi) = A \sin\roundb{\omega_m \xi} \,,
    \qquad
    \gamma_m(\xi) = - A \dfrac{\alpha p_m^\pm}{\alpha\zeta + p_m^\pm} \sin\roundb{\omega_m \xi} \,,
\end{equation}
where  $A$ is an arbitrary amplitude.

\paragraph{The emergence of a strongly discontinuous bifurcation mode.}
In addition to the twin sequences of bifurcation loads $p_m^{\pm}$, the bifurcation of the cantilever rod also occurs at a transition  load $p_0$ which results coincident with $p^*$, \cref{eq:ContAccumulationLoad},
\begin{equation}\label{eq:ContCantilevLocalLoad}
    p_0 = p^*,
\end{equation}
consistent with the finding from  the analysis of  the cantilever discrete chain, \cref{eq:DiscCantilevLocalLoad}. As a matter of fact, when $p=p_0=p^*$, the algebraic Eq.~(\ref{eq:ContCantilevLinearEq})$_2$ cannot anymore be solved for $\gamma(\xi)$, but merely reduces to the condition $\theta(\xi)=0$.

By using the dimensionless coordinate  $\xi = s/L$ and enforcing the linear approximation underlying the bifurcation analysis, the solvability condition \eqref{eq:ContApplicationExactMM} reduces to
\begin{equation}\label{eq:ContCantilvBucklingMM}
     \alpha\zeta \int_0^1 \gamma(\xi) \,\dxi =  \theta'(0) \,,
\end{equation}
which suggests that the solution is to be sought within the framework of generalized functions in the form $\gamma(\xi) = c \,\delta(\xi)$, where $\delta(\xi)$ is the Dirac delta function and $c$ is an arbitrary amplitude, implying $\theta'(0) = c\, \alpha\zeta$. In this circumstance, \cref{eq:ContCantilevLinearEq}\ped{1} leads to a representation of $\theta'$ in terms of Heaviside step function $H$, namely, 
$\theta'(\xi)=c\,\alpha\zeta [1-H(\xi)]$, so that $\theta(\xi)=c\, \alpha\zeta\, \xi [1-H(\xi)]$ vanishes for $\xi \in [0,1]$.

The bifurcation mode just found corresponds to the rod remaining straight under the axial load, but rigidly displaced in the transverse direction of an arbitrary amount. Therefore, both the displacement and the bending moment suffer a jump at the clamp, where the shear angle $\gamma(\xi)$ displays a stronger singularity. This behaviour is shown in Fig.~\ref{fig:cantilever} in the inset labelled \lq Transition mode'.

\subsubsection{Simply supported configuration for the equivalent rod}

The bending moment at both ends of the rod, Fig.~\ref{fig:BucklingSchemes}b (lower part), is null, so that the boundary conditions on the function  $\theta(s)$, Eqs.~\eqref{eq:ContRodEquilBC}, read as
\begin{equation}\label{eq:ContSimplySuppBoundaryCond}
    \theta'(0) = 0 \,,
    \qquad
    \theta'(L) = 0 \,.
\end{equation}
complementing the differential-algebraic problem \eqref{eq:ContApplicationExactEq}.

Using again the dimensionless parameters, \cref{eq:ForceStiffnessNormalization}, and the normalized coordinate $\xi \in \squareb{0,1}$, the linearized equations \eqref{eq:ContCantilevLinearEq} are again obtained.

When $p \neq p^*$, \cref{eq:ContAccumulationLoad}, the function $\gamma(\xi)$ is again given by \cref{eq:ContCantilevGamma2Theta} and the homogeneous differential equation \eqref{eq:ContCantilevDiffEq} still applies, with its general solution, \cref{eq:ContCalilevGeneralSol}. The boundary conditions \cref{eq:ContSimplySuppBoundaryCond} provide $A = 0$ and $\omega \sin{\omega} = 0$. The latter condition is satisfied when 
\begin{equation}\label{eq:ContSimplySuppEulerCrit}
    \omega_m^2 = m^2 \pi^2 \,,
        \qquad m = 1, 2, \ldots,
\end{equation}
which can be used in \cref{eq:ContCantilevCritLoads} to obtain the bifurcation loads, while the relevant modes become
\begin{equation}\label{eq:ContSimplySuppBifurcationShapes}
    \theta_m(\xi) =  B \cos\roundb{\omega_m \xi} \,,
    \qquad
    \gamma_m(\xi) = - B \dfrac{\alpha p_m^\pm}{\alpha\zeta + p_m^\pm} \cos\roundb{\omega_m \xi} \,,
\end{equation}
being $B$ an arbitrary amplitude.

Differently from the cantilever scheme, for the simply supported rod the condition $\omega = 0$ provides an additional bifurcation load. In fact, recalling the definition \eqref{eq:DiscCanilevQuadEigen} of $\omega^2$, the transition load $p_0$ can be derived as in Eq. \eqref{eq:DiscSimplySuppLocalLoad}, 
valid also for the continuous rod, and the corresponding mode is
\begin{equation}\label{eq:ContSimplySuppLocalShape}
    \theta_m(\xi) = -\gamma_m(\xi)=  B \,.
\end{equation}

In the degenerate case $p = p^*$, \cref{eq:ContCantilevLinearEq}\ped{2} provides $\theta(\xi) = 0$, and, using the boundary conditions \eqref{eq:ContSimplySuppBoundaryCond} in \cref{eq:ContApplicationExactMM}, the solvability condition for the linearized buckling analysis becomes
\begin{equation}
     \alpha\zeta \int_0^1 \gamma(\xi) \,\dxi = 0 \,,
\end{equation}
an equation that is satisfied by the null function $\gamma(\xi) = 0$, so that, differently from the case of the cantilever rod,  \cref{eq:ContCantilvBucklingMM},
a discontinuity in $\gamma(\xi)$ does not occur.

\subsection{Patterns of bifurcations for the discrete chain and the  equivalent rod}

The results obtained in the previous Sections \ref{discretino} and \ref{equivalentino} for the discrete chain are compared with analogous results relative to the equivalent shearable rod. In particular,  bifurcation loads, Eqs. \eqref{eq:DiscCantilevCritLoads}, \eqref{eq:DiscCantilevLocalLoad}, \eqref{eq:DiscSimplySuppLocalLoad}, \eqref{eq:ContCantilevCritLoads}, and \eqref{eq:ContCantilevLocalLoad}, and  bifurcation modes, Eqs. \eqref{eq:DiscCantileverLocalShape}, \eqref{eq:DiscSimplySuppLocalShape}, \eqref{eq:ContCantilevBifurcationShapes}, \eqref{eq:ContSimplySuppBifurcationShapes}, and \eqref{eq:ContSimplySuppLocalShape}, are investigated for the cantilever and the simply supported equivalent rod schemes.

\subsubsection{The bifurcation loads and modes}

The bifurcations of the cantilever chain and the simply supported chain are characterized in both cases by the twin sequences of bifurcation mechanisms, Eqs.~\eqref{eq:DiscCantilevCritLoads}, separated by a transition mode, Eqs.~\eqref{eq:DiscCantilevLocalLoad} and \eqref{eq:DiscSimplySuppLocalLoad}.
The twin modes for the discrete systems correspond to $p^+_m(n)$ and $p^-_m(n)$ and are characterized by the same shape of the nodes' interpolation obtained through different contributions  of the angles $\theta^i$ (representing the rotation of the chain elements) and  the angles $\beta^i$ (characterizing the micromechanics connected with the inclination of the four-bar linkage), Fig.~\ref{fig:DiscreteLinkageKinem}. 
The same feature also characterizes the bifurcation modes corresponding to $p^+_m$ and $p^-_m$ of the continuous system, Eqs.~\eqref{eq:ContCantilevCritLoads}, so that the same deformed shape of the rod axis is obtained for both twin modes by considering different angles ratios $\theta(s)/\gamma(s)$, Eqs.~\eqref{eq:ContCantilevBifurcationShapes} and \eqref{eq:ContSimplySuppBifurcationShapes}. 
It follows that the overall shape of the bifurcation modes is the same, but the actual configuration is obtained via different contributions of the rotation and the shear angles.

\begin{figure}[ht!]
    \centering
    \includegraphics[width=170mm]{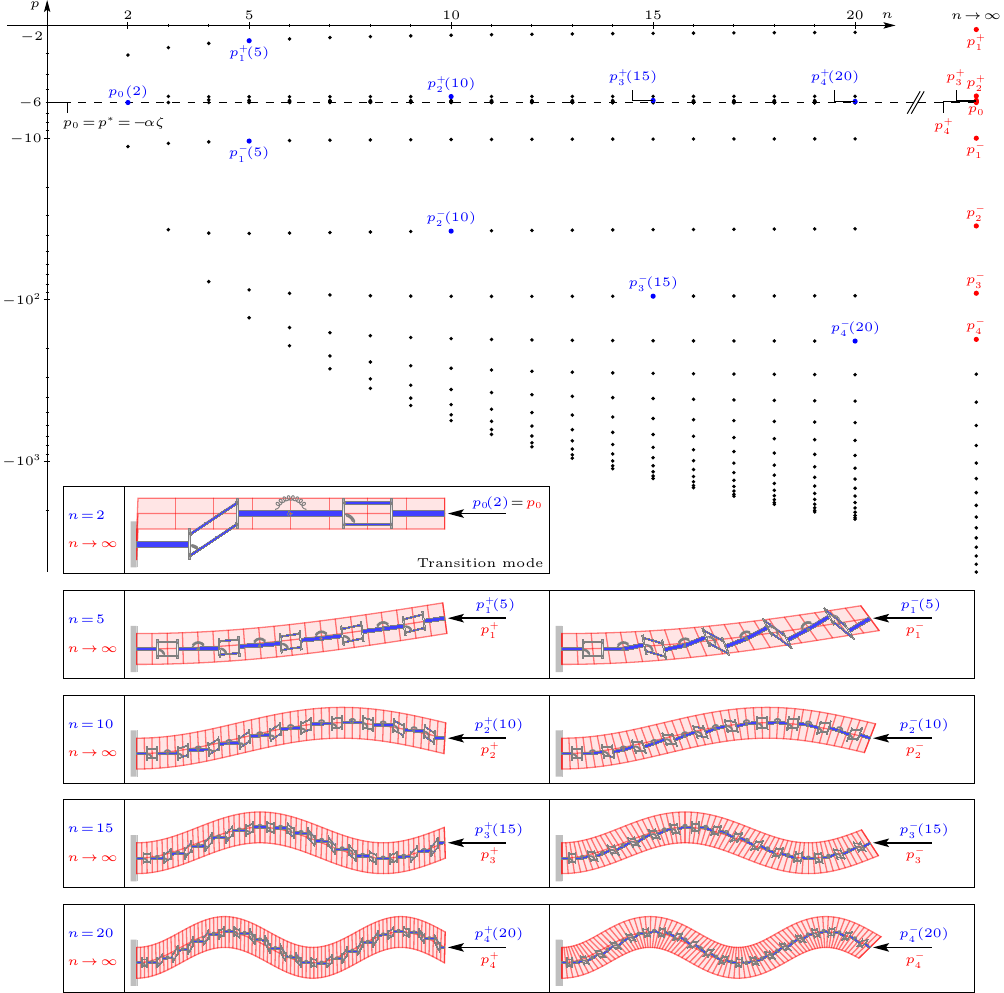}
    \caption{Bifurcation loads and modes for the cantilever chain and for the equivalent elastic rod ($\alpha = 0.3 \,, \zeta = 20$). The discrete systems are defined by a number of elements $n$ ranging between 2 and 20. The twin sequences of bifurcation loads \lq $+$' and \lq $-$' are separated by the \lq transition' load $p_0 = p^*$ and are reported for the discrete systems (marked blue), corresponding to $n$ values, and for the continuous system (marked red). 
    The twin $m$-th critical modes for the continuous system are reported superimposed to chains with increasing number $n$ of elements. Note that the transition mode involves a jump in displacement at the clamp.
    }
    \label{fig:cantilever}
\end{figure}

\begin{figure}[ht!]
    \centering
    \includegraphics[width=170mm]{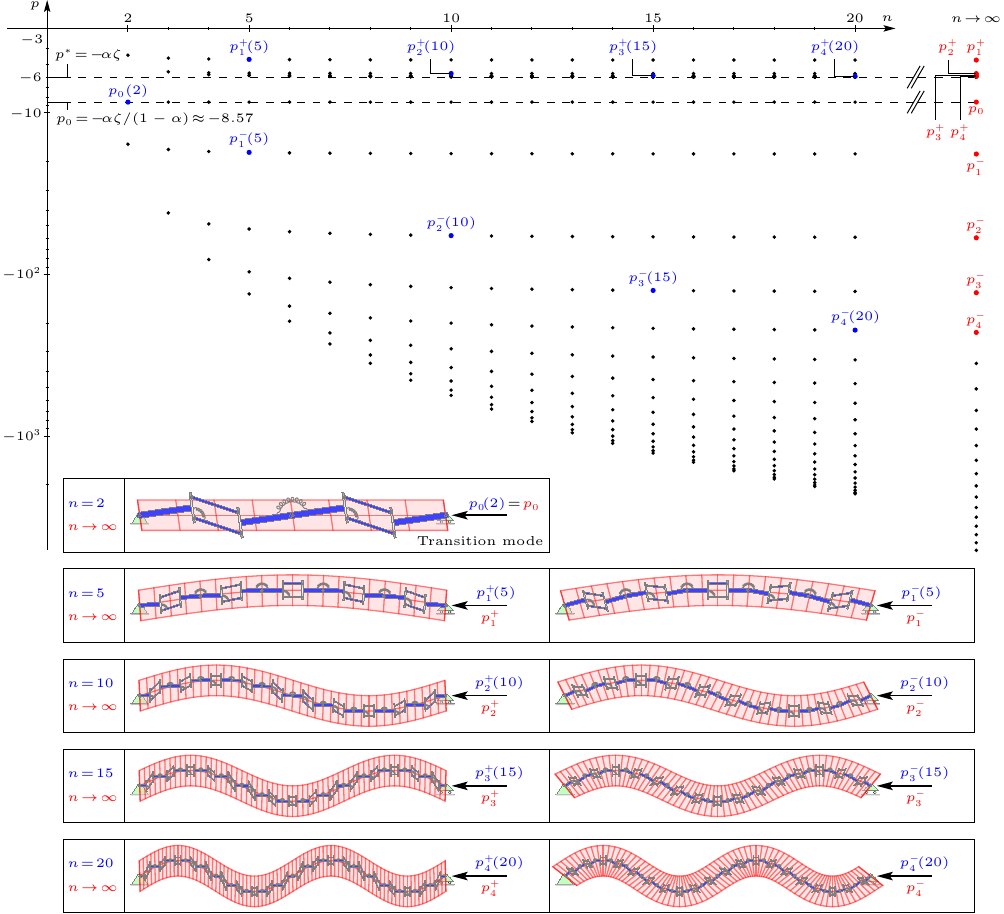}
    \caption{As for Fig.~\ref{fig:cantilever}, except that the rod is simply supported. 
    In this case $p^*>p_0$ and the transition mode has a  \lq bookshelf-like' shape.
    }
    \label{fig:SimplySuppBifurcation}
\end{figure}

Figs.~\ref{fig:cantilever} and \ref{fig:SimplySuppBifurcation} 
report bifurcation loads and modes for the discrete and continuous systems. The former figure is dedicated to the cantilever scheme and the latter to the simply supported one. 
Both figures pertain to a system with a low  stiffness ratio, $\zeta = 20$, and a value $\alpha = 0.3$ of the microstructural parameter, 
providing $p^*=-6$. Parameters $\zeta$ and $\alpha$ have been selected as representative of a structure in which the shear deformation plays an important role.

A set of nineteen different discrete systems are considered, corresponding to $n$ varying between 2 and 20, with the bifurcation mode  number $m$
satisfying $m \leq n-1$. 
The continuous and discrete structures exhibit the twin sequences of bifurcation loads, separated by the transition load
$p_0$, which is independent of $n$ and therefore appears as a constant value in the figures. 
Note that for the continuous model (identified by $n\to\infty$ in the figures) the bifurcation mode number $m$ ranges from 1 to $+\infty$ and the corresponding modes are sketched in red. 

In the continuous model for both cantilever and simply supported conditions,  the first sequence of bifurcation loads initiates from $p^+_1$ and accumulates towards $p^*$, \cref{eq:ContAccumulationLoad}, namely, 
\begin{equation}
\lim_{m\to\infty}p_m^+=p^*,
\end{equation}
while the second sequence initiates from $p^-_1 < p_0$ and grows to infinity, $p^-_\infty \to -\infty$. For the cantilever rod the accumulation load coincides with the transition load,  $p_0=p^*$, Fig.~\ref{fig:cantilever}, while these two loads remain different for the simply supported rod and verify $p_0 < p^*$, Fig.~\ref{fig:SimplySuppBifurcation}.

The bifurcation modes characteristic of the twin sequences are reported in the figures for $p_m^+(n)$ on the left and for $p_m^-(n)$ on the right. 
The twin modes display the same deformed rod axis but the rotation of the cross-sections is different, 
so that each bifurcation mode of the chain (with a given number $n$ of discrete elements) corresponds to a unique bifurcation mode of the continuous rod. 
The modes belonging to the 
 sequence \lq $+$', are characterized by the angles $\theta$ and $\gamma$ sharing the same sign, while these signs become opposite for the modes belonging to the second sequence. 

The transition mode exhibits a peculiar shape, because a jump in displacement for the cantilever rod emerges at the clamp (Fig.~\ref{fig:cantilever}, $n=2$, $p_0(2)=p_0$), while 
a \lq bookshelf-like' bifurcation mode 
(Fig.~\ref{fig:SimplySuppBifurcation}, $n=2$, $p_0(2)=p_0$)
characterizes the simply supported scheme. These bifurcation modes,   observed in the equivalent rod, closely resemble their discrete counterparts (drawn superimposed on the deformed rod).

A comparison between the numerical values of bifurcation loads for discrete and continuous systems shows that the critical loads characterizing the first sequence and 
the transition mode are captured with an excellent  approximation already with a chain of 5 elements only. 
This approximation becomes worse for critical loads smaller than that corresponding to the transition mode.

\subsection{The special case of the Engesser rod}

It has been shown in Section \ref{sec:EngesserGeneral} that our model of  equivalent rod specializes to the Engesser rod in the limit $\alpha \to 1$, so that the bifurcation analysis for the  Engesser rod can be performed by specializing the results obtained from Eqs.~\eqref{eq:ContCantilevLinearEq}.

With reference to the cantilever and the simply supported rods shown in Fig.~\ref{fig:BucklingSchemes}, in the limit for $\alpha \to 1$, the second sequence of bifurcation loads, \cref{eq:ContCantilevCritLoads}, diverges, $p_m^- \to -\infty$, while the first sequence can be expressed as
\begin{equation}\label{eq:EngesserBifurcationLoads}
    p_m = -\dfrac{\zeta \omega_m^2}{\zeta + \omega_m^2}  \,, \quad m = 1, 2, \ldots \,,
\end{equation}
where $\omega_m^2$ is given by \cref{eq:ContCantilevEulerCrit} and \cref{eq:ContSimplySuppEulerCrit} for the cantilever and the simply supported rod, respectively. Moreover, the relevant bifurcation modes can be obtained by specializing Eqs.~\eqref{eq:ContCantilevBifurcationShapes} and \eqref{eq:ContSimplySuppBifurcationShapes}, whence the rotation angle of the tangent to the rod axis, for the two schemes, becomes
\begin{equation}\label{eq:EngesserBifurcationShapes}
    \varphi_m(\xi) = A \dfrac{\zeta}{\zeta + p_m} \sin\roundb{\omega_m \xi} \,,
    \qquad
    \varphi_m(\xi) = B \dfrac{\zeta}{\zeta + p_m} \cos\roundb{\omega_m \xi} \,,
\end{equation}
where the two expressions refer to the cantilever and the simply supported rod, respectively, and $A$ and $B$ are arbitrary amplitude coefficients.

The transition load for the cantilever structure, \cref{eq:ContCantilevLocalLoad}, specializes to $p_0 = - \zeta$, so that, even for the Engesser model, a cantilever rod exhibits a local buckling mechanism, with $\gamma(\xi) = c\, \delta(\xi)$, $\theta(\xi) = 0$ and $\varphi(\xi) = c\, \delta(\xi)$, where $\delta(\xi)$ is the Dirac delta function and $c$ is an arbitrary amplitude. On the contrary, the transition load $p_0$ of the simply supported rod, \cref{eq:DiscSimplySuppLocalLoad}, diverges to $-\infty$ in the limit $\alpha \to 1$, so that the \lq bookshelf-like' bifurcation mode does not occur for the Engesser rod model. Furthermore, the sequence $p_m$ of bifurcation loads, \cref{eq:EngesserBifurcationLoads}, accumulates to the value $p^* = -\zeta$,  \cref{eq:ContAccumulationLoad}, which represents a minimum for the cantilever rod and an infimum for the simply supported scheme.

The bifurcation analysis can also be conducted from the equation governing the mechanics of the Engesser rod, \cref{eq:EngesserSingleEq}, with $\BF_L = P \Be_1$ and null distributed load $\bar{\Bq}(s)$ and couple $\bar{m}(s)$. For this load conditions, the equation governing the Engesser rod model becomes
\begin{equation}\label{eq:EngesserApplicationExact}
    EI \squaree{\varphi(s) + \dfrac{P}{GA\ped{s}} \sin{\varphi(s)}}''
    - P \, \sin{\varphi(s)} = 0 \,,
\end{equation}
which coincides with the equation derived for the so-called \lq Timoshenko's approach' \cite{atanackovic} and the \lq Engesser elastica' \cite{Kocsis}. The boundary conditions, Eqs.~\eqref{eq:EngesserBoundaryC0} and \eqref{eq:EngesserBoundaryCL}, become
\begin{equation}\label{eq:EngesserCantileverBCExact}
    \varphi(0) + \dfrac{P}{GA\ped{s}} \sin{\varphi(0)} = 0 \,,
    \qquad
    \rounde{1 + \dfrac{P}{GA\ped{s}} \cos{\varphi(L)}}\varphi'(L) = 0 \,,
\end{equation}
for the cantilever rod, and
\begin{equation}\label{eq:EngesserSimplysupBCExact}
    \rounde{1 + \dfrac{P}{GA\ped{s}} \cos{\varphi(0)}}\varphi'(0) = 0 \,,
    \qquad
    \rounde{1 + \dfrac{P}{GA\ped{s}} \cos{\varphi(L)}}\varphi'(L) = 0 \, ,
\end{equation}
for the simply supported scheme.

Using the dimensionless parameters defined by Eqs.~\eqref{eq:ForceStiffnessNormalization}, the dimensionless coordinate $\xi = s/L$, and applying the linear approximations $\sin{\varphi(\xi)} \approx \varphi(\xi)$ and $\cos{\varphi(\xi)} \approx 1$, \cref{eq:EngesserApplicationExact} becomes
\begin{equation}\label{eq:EngesserApplicationLinearized}
    \rounde{1 + \dfrac{p}{\zeta}}\varphi''(\xi)
    - p \, \varphi(\xi) = 0 \,,
\end{equation}
and the boundary conditions, Eqs.~\eqref{eq:EngesserCantileverBCExact} and \eqref{eq:EngesserSimplysupBCExact}, reduce to
\begin{equation}\label{eq:EngesserLinearBC}
    \underbrace{\rounde{1 + \dfrac{p}{\zeta}}\varphi(0) = 0 \,,
    \quad
    \rounde{1 + \dfrac{p}{\zeta}}\varphi'(1) = 0 \,,
    }_\text{Cantilever rod}
    \qquad
    \underbrace{
    \rounde{1 + \dfrac{p}{\zeta}}\varphi'(0) = 0 \,,
    \quad
    \rounde{1 + \dfrac{p}{\zeta}}\varphi'(1) = 0 \,.
    }_\text{Simply supported rod}
\end{equation}

When $p \neq p^*= -\zeta$, it is possible to set $\omega^2 = - \zeta p/\roundb{\zeta + p}$, so that the differential equation \eqref{eq:EngesserApplicationLinearized} can be expressed as \cref{eq:ContCantilevDiffEq} with $\varphi(\xi)$ replacing  $\theta(\xi)$. Moreover, the boundary conditions, \cref{eq:EngesserLinearBC}, simplify to \big($\varphi(0) = 0$, $\varphi'(1) = 0$\big) for the cantilever rod, and ($\varphi'(0) = 0$, $\varphi'(1) = 0$) for the simply supported scheme, which are analogous to Eqs.~\eqref{eq:ContCantilevBoundaryCond} and \eqref{eq:ContSimplySuppBoundaryCond}.

Therefore, the values $\omega_m^2$ given by Eqs.~(\ref{eq:ContCantilevEulerCrit}) and (\ref{eq:ContSimplySuppEulerCrit})  also apply to the buckling of the Engesser rod, the bifurcation loads are given by \cref{eq:EngesserBifurcationLoads} and the bifurcation modes coincide with Eqs.~\eqref{eq:EngesserBifurcationShapes}.

Note that when $p =p^*= -\zeta$, corresponding to a local buckling mechanism for the cantilever rod, \cref{eq:EngesserApplicationLinearized} reduces to $\varphi(\xi) = 0$ and the boundary conditions in terms of $\varphi(\xi)$, Eqs.~\eqref{eq:EngesserLinearBC}, do not apply. Consequently,  \cref{eq:EngesserApplicationLinearized} becomes useless, because its validity is limited to the bifurcation modes described by continuous functions. On the other hand, for a comprehensive analysis of the Engesser rod, a differential-algebraic system of equations in the form of  Eqs.~\eqref{eq:ContCantilevLinearEq} is required.

\section{Postcritical response of the simply supported rod showing the emergence and growth of folding}

The examination of the postcritical response of the  simply supported rod under compression requires 
the development of numerical codes to solve the nonlinear equations 
governing the quasi-static response of the discrete chain, Eqs.~\eqref{eq:DiscApplicationExactEq}, and of its equivalent rod, Eqs.~\eqref{eq:ContApplicationExactEq}.

\subsection{The continuous rod and the numerical solution of its deformation under load}

 The numerical solution of the differential-algebraic equations \eqref{eq:ContApplicationExactEq} is performed by considering the equivalent system of nonlinear differential equations, expressed in terms of dimensionless parameters \eqref{eq:ForceStiffnessNormalization} and of the  coordinate $\xi = s/L$, as
\begin{equation}\label{eq:ContPostBucklingEqns}
\begin{array}{l}
    \theta''(\xi)  = p \, \roundb{1-\alpha} \sin{\theta(\xi)}  - \alpha\zeta \, \gamma(\xi) \,, \\[2ex]
    \left[\alpha\zeta + p\cos\roundc{\theta(\xi) + \gamma(\xi)/\alpha}\right]\gamma'(\xi) = - \alpha p\, \cos\roundc{\theta(\xi) + \gamma(\xi)/\alpha} \, \theta'(\xi) \,,
\end{array}
\qquad
\xi\in(0,1) \,,
\end{equation}
to be complemented with the dimensionless versions of Eqs.~\eqref{eq:ContApplicationExactEq}\ped{2} and \eqref{eq:ContSimplySuppBoundaryCond}. The nonlinear analysis is performed by decreasing the dimensionless load $p<0$ through discrete load steps  $\Delta p < 0$ (of amplitude varying with the incremental stiffness) and  the system \eqref{eq:ContPostBucklingEqns} is integrated at each step using the function \lq \texttt{bvp4c}' of Matlab  \cite{Matlab}. 

\begin{figure}[ht!]
    \centering
    \includegraphics[width=170mm]{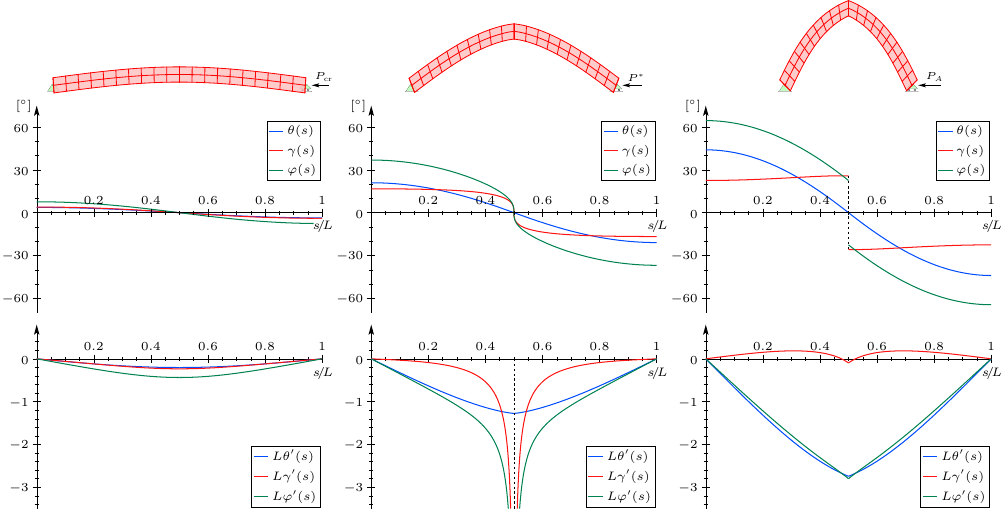}
    \caption{Upper part: Three stages during the development of folding in the postcritical behaviour of the equivalent rod. 
    Central part: Rotation of the cross-section $\theta$, of the tangent to the axis $\varphi$, and shear angle $\gamma$. 
    Lower part: Curvature of the rod axis $\varphi'$, and derivatives of angles $\theta$ and $\gamma$. 
    The graphs refer to three levels of negative load $P$, with increasing modulus from left to right: at first bifurcation $P\ped{cr}$ (left), at the emergence of folding for the  load $P^*$ (centre, displaying infinite value for the mid-span curvature $\varphi'(L/2)$), and when folding is already developed $P_A$ at $u_1(L)/L = -0.5$ (right, displaying at the mid-span the formation of a cusp in the curvature $\varphi'(1/2)$ and a jump in the shear angle $\gamma(1/2)$). 
    }
    \label{fig:ContDeformFunctions}
\end{figure}

Note that, because of the symmetry of the rod configuration, the rotation $\theta(\xi)$ vanishes at the midpoint $\xi=1/2$ and, by virtue of the algebraic equation \eqref{eq:ContApplicationExactEq}\ped{2}, the shear angle $\gamma(\xi)$ is also null at this point. This symmetry condition leads to a problem during the integration of the system (\ref{eq:ContPostBucklingEqns}), when the dimensionless load $p$ reaches the value $p^*$, \cref{eq:ContAccumulationLoad}. In fact, for this value of the load (representing the limit value of the first sequence $p_m^+$ of buckling loads), the coefficient of $\gamma'(\xi)$ in \cref{eq:ContPostBucklingEqns}\ped{2} vanishes at $\xi = 1/2$. Consequently,  function $\gamma(\xi)$ is not defined at $\xi = 1/2$ when $p=p^*$, but approaches zero with a vertical tangent from both sides of this point, while the symmetry condition still applies as a limit (Fig.~\ref{fig:ContDeformFunctions}, centre), 
\begin{equation}
    \mylim{|\varepsilon|\to 0}{\, \gamma\left(1/2\pm|\varepsilon|\right)} =0 \,,
    \qquad
    \mylim{|\varepsilon|\to 0}{\, \gamma'\left(1/2\pm|\varepsilon|\right)} = -\infty, \qquad \text{for}\enspace p=p^*  \,.
\end{equation}

The singularity in $\gamma(\xi)$ and $\gamma'(\xi)$ at $\xi = 1/2$ also affects the rotation angle $\varphi(s)$ of the tangent to the rod axis and its curvature, $\varphi'(s)$. Therefore, Eqs.~\eqref{eq:ContAxisRotationAngle} and \eqref{eq:ContAxisCurvature} 
lead to 
\begin{equation}
    \mylim{|\varepsilon|\to 0}{\, \varphi\left(1/2\pm|\varepsilon|\right)} = 0 \,,
    \qquad
    \mylim{|\varepsilon|\to 0}{\,\varphi'\left(1/2\pm|\varepsilon|\right)}  = -\infty, \qquad \text{for}\enspace p=p^* \,,
\end{equation}
whence the curvature $\varphi'(s)=\varphi'(\xi)/L$ diverges to infinity
and marks the emergence of folding at the mid-span of the rod.

 When the (negative) force $P$ is further increased in its modulus ($P<P^*$), a jump occurs at the mid-span ($\xi=1/2$) in the shear angle $\gamma$ and consequently  
 in the rotation angle $\varphi$ of the tangent to the rod axis, while their derivatives ($\gamma'$ and $\varphi'$) remain finite and continuous (Fig.~\ref{fig:ContDeformFunctions}, right). More specifically, due to symmetry, the following properties hold
 \begin{equation}\begin{array}{lll}
      &\gamma\left(1/2^+\right) =- \gamma\left(1/2^-\right)\neq 0 \,,
    \qquad
    \gamma'\left(1/2^+\right)  =  \gamma'\left(1/2^-\right)\neq \pm\infty \,,  \\[3ex]
      &  \varphi\left(1/2^+\right) =-  \varphi\left(1/2^-\right)\neq 0 \,,
    \qquad
    \varphi'\left(1/2^+\right)  = \varphi'\left(1/2^-\right)\neq\pm \infty \,, 
 \end{array}  
 \qquad \text{for}\enspace p<p^* \,.
\end{equation}
It is also noted that, because symmetry is maintained during the deformation path, although  $\gamma$ is discontinuous at the mid-span, $\lambda$ remains continuous at that point, 
\begin{equation}
    \lambda\left(1/2^+\right)=\lambda\left(1/2^-\right) \,.
\end{equation}
 
As a consequence of the singularity of functions $\gamma(\xi)$ and $\gamma'(\xi)$ in \cref{eq:ContPostBucklingEqns}, the Runge-Kutta scheme implemented in the numerical solver fails to converge when applied to the entire domain $[0, 1]$. The strategy adopted to overcome this problem is to release the continuity of the angle 
$\theta$ 
at mid-span by disconnecting the left and right halves of the rod through 
the insertion of 
an elastic hinge of stiffness $K_0$ (made dimensionless as $\kappa_0 = K_0 L/(EI)$). The adopted numerical strategy is similar to 
the introduction of finite elements with embedded strong discontinuities in the shear band analysis of continua, see among many others \cite{armero}.

Notice that the insertion of an elastic hinge at mid-span of the equivalent rod modifies the structural setup and thus the bifurcation results. In fact,  when a simply supported rod contains an elastic hinge at mid-span, 
the solution of the linearized differential equation \eqref{eq:ContCantilevDiffEq} 
becomes
\begin{equation}
    \theta(\xi) = \left\{
        \begin{array}{lr}
              A_1 \sin\roundb{\omega \xi} + B_1 \cos\roundb{\omega \xi} \,, &  \text{if}\enspace \xi \in \left[0, 1/2\right) \,,\\[2ex]
              A_2 \sin\roundb{\omega \xi} + B_2 \cos\roundb{\omega \xi} \,, &  \text{if}\enspace \xi \in \left(1/2, 1\right] \,,
        \end{array}
        \right.
\end{equation}
and the boundary conditions $\theta'(0) = \theta'(1) = 0$ are supplemented by the constraints holding across the hinge, 
\begin{equation}
\theta'(1/2^-) = \theta'(1/2^+) \,,
\qquad
\theta(1/2^-) - \theta(1/2^+) = \theta'(1/2^-)/\kappa_0 \,.
\end{equation}
Non-trivial solutions occur when 
\begin{equation}\label{eq:ContSimplySupBucklingK0}
    \sin{\omega} = \dfrac{\omega\roundb{1-\cos{\omega}}}{2\kappa_0}  \,,
\end{equation}
and for a value $\omega=\omega_m$, identifying the $m$--th bifurcation mode, twin bifurcation loads $p_m^\pm$ are obtained from  \cref{eq:ContCantilevCritLoads}. 
Notice that $\omega_1$, and thus the critical load $p_1^+$, is lower than that obtained in the absence of the hinge and approaches the value corresponding to the intact rod when  $\kappa_0$ increases to infinity, leading \cref{eq:ContSimplySupBucklingK0} to coincide with the condition $\sin{\omega} = 0$, pertaining to the intact rod.

The post-buckling behaviour is now obtained by numerically integrating Eqs.~\eqref{eq:ContPostBucklingEqns} on two symmetric subdomains separately. The solution depends on $\kappa_0$, but it is found to converge to a fixed response when the stiffness is increased at infinity, $\kappa_0 \to \infty$, a condition which restores the continuity of the rotation $\theta(s)$, but allows the discontinuity of the shear angle $\gamma(s)$ to persist. 
The convergence of the solution at increasing $\kappa_0$ demonstrates that the same solution can be obtained by simply dividing the integration domain into two halves ($\xi\in[0,1/2)$ and $\xi\in(1/2,1]$), constrained only through the continuity condition in $\theta$ and 
$\theta'$ at $\xi=L/2$,
\begin{equation}
\theta'(1/2^-) = \theta'(1/2^+)\,,
\qquad
\theta(1/2^-) = \theta(1/2^+) \,.
\end{equation}

The process of folding nucleation and growth is illustrated in Fig.~\ref{fig:ContDeformFunctions}, where three snapshots of the 
variation of functions $\theta$, $\gamma$, $\varphi$ and their derivatives  along the rod axis is reported. The graphs are complemented in the upper part with the deformed shape of the equivalent rod at the three different values of negative loads $P$: (i.) the critical buckling load $P\ped{cr}$, (ii.) the load  $P^*=p^*EI/L^2$ providing the emergence of folding, and (iii.) at a later stage 
$P_A$, corresponding to $u_1(L) = -L/2$. 

Folding emerges at the load $p = p^*$ as a point-wise singularity of $\gamma(s)$, and also of $\varphi(s)$, corresponding to infinite curvature at the midpoint, Fig.~\ref{fig:ContDeformFunctions} (central column). When the load is further decreased, $p < p^*$, the shear angle $\gamma(s)$ develops a jump discontinuity growing with the load, Fig.~\ref{fig:ContDeformFunctions} (right column).

\begin{figure}[ht!]
    \centering
    \includegraphics[width=160mm]{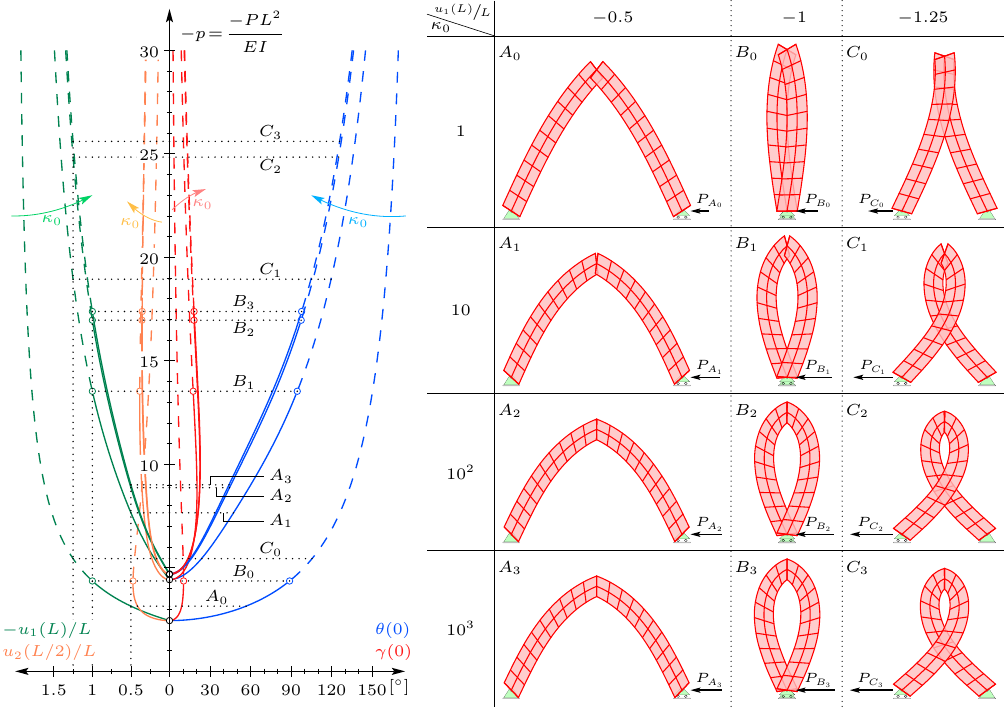}
    \caption{Post-buckling and folding development of the simply supported equivalent rod: load vs end angles $\theta(0)$, 
    $\gamma(0)$ and displacements $u_1(L)$ and $u_2(L/2)$ (left); deformed configurations at different loads, points $A_j$, $B_j$, and $C_j$ ($j=\{0,1,2,3\}$, respectively corresponding to $\kappa_0=\{1, 10, 10^2, 10^3\}$) (right). 
    A discontinuity in the curvature of the rod axis at the midpoint is modelled by introducing an elastic hinge with stiffness $\kappa_0 = K_0 L/(EI)$. Four values of rotational stiffness are reported, showing that the solution tends to a fixed value of folding when the stiffness increases to infinity. Folding leads the rod to assume the shape of a pointed arch. See also the video file in the electronic supplementary material.
    }
    \label{fig:SimplysupContPath}
\end{figure}

For the equivalent rod model, 
Fig.~\ref{fig:SimplysupContPath} depicts  
the applied force as a function of the 
cross-section rotation $\theta(0)$, shear angle $\gamma(0)$, and  normalized displacements in both the $\Be_1$ 
and in the $\Be_2$ 
directions of the endpoint $u_1(L)$ and of the midpoint $u_2(L/2)$, respectively.
The rod is disconnected at mid-span through the inclusion of an elastic hinge of dimensionless stiffness $\kappa_0$, so that the four values $\kappa_0=\{1, 10, 10^2, 10^3\}$ are reported. 
It is worth mentioning that the four curves tend to accumulate on that corresponding to $\kappa_0=10^3$ and that higher values of $\kappa_0$ (not reported for conciseness) result in superimposed curves. This provides evidence to the fact that the continuous solution converges to folding, a feature clearly emerging from the bifurcation modes reported on the right of the graphs, where the rows refer to the different  stiffnesses of the rotational spring. 

Note that Fig.~\ref{fig:SimplysupContPath} refers to the first mode $m=1$ (higher mode bifurcations are not explored) and points marked on the graphs as $A_j$, $B_j$, and $C_j$ ($j=\{0,1,2,3\}$, respectively corresponding to $\kappa_0=\{1, 10, 10^2, 10^3\}$) correspond to the deformations reported on the right. The continuous curves denote stable behaviour, while dashed lines indicate an unstable path. The latter always initiates when the two supports coincide, as illustrated in the central column of the figure. The folding developing at mid-span and making the deformed shape similar to a pointed arch is displayed (left column of the figure).

\paragraph{The Engesser rod.} The Engesser model shares with our equivalent one  the development of  folding in the post-buckling behaviour of a simply supported configuration. This feature, never noticed before, becomes evident from Eqs.~\eqref{eq:ContPostBucklingEqns} which hold true also in the limit condition $\alpha \to 1$, so that a singularity at $\xi = 1/2$ emerges for  the functions $\gamma(\xi)$ and $\gamma'(\xi)$ at $p = p^* = -\zeta$. The singularity also affects the rotation angle $\varphi(\xi)$ of the tangent to the axis and the curvature $\varphi'(\xi)$, the latter marking the emergence of folding.

\subsection{Numerical integration of the equilibrium equations for the discrete system}

\begin{figure}[ht!]
    \centering
    \includegraphics[width=170mm]{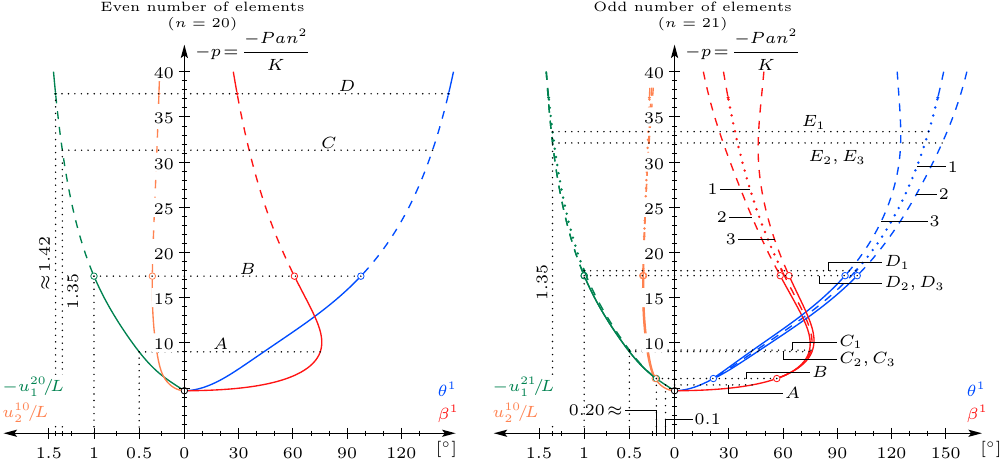}
    \caption{Post-buckling of simply supported chains with even ($n = 20$)  and odd ($n = 21$) number of elements. 
    Equilibrium paths (force $-P$ vs kinematic parameters $\theta^1$, $\beta^1$, $u_1$, and $u_2$) for the even (left) and the odd (right) system. Stable (unstable) paths are denoted with continuous (dashed) lines and dots indicate bifurcation points. Loss of stability occurs at bifurcation when the two supports coincide, $B$, and restabilization, $D$, is achieved later for the even system. The odd system shows a secondary, $B$ and three tertiary, $D_j$ ($j=1,..,3$), bifurcations.
}
    \label{fig:SimplysupDiscGRAPH}
\end{figure}

\begin{figure}[ht!]
    \centering
    \includegraphics[width=170mm]{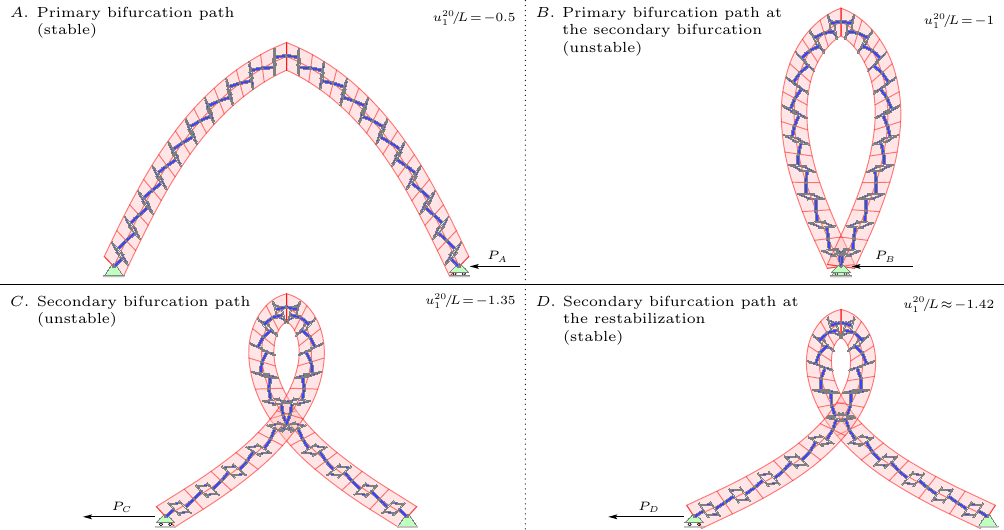}
    \caption{Deformed configurations during the post-buckling of a simply supported chain with an even number, $n = 20$, of  elements. The deformed configurations are superimposed  to the response of the  homogenized rod, highlighting the validity of the homogenization scheme and the emerging of folding at the midpoint. Points $A$, $B$, $C$, and $D$ correspond to Fig.~\ref{fig:SimplysupDiscGRAPH} on the left. See also the video file in the electronic supplementary material.}
    \label{fig:SimplysupDiscConfEVEN}
\end{figure}

\begin{figure}[ht!]
    \centering
    \includegraphics[width=170mm]{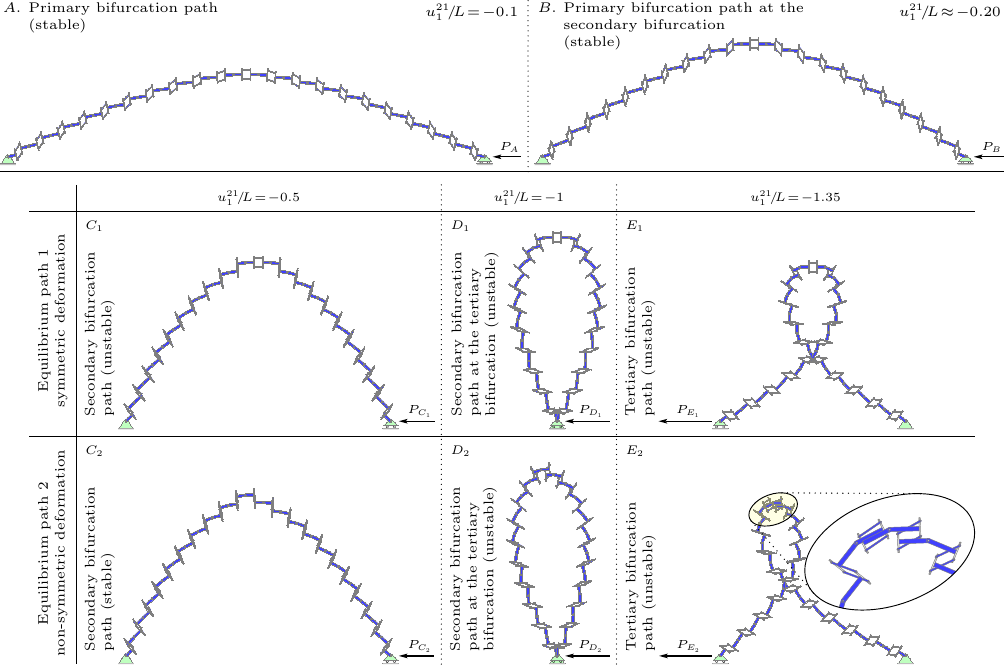}
    \caption{
    Deformed configurations during the post-buckling of a simply supported chain with an odd number, $n = 21$, of  elements.
    Secondary and tertiary bifurcations are found. In particular, the secondary (stable) bifurcation path mimics the development of folding with a strong distortion of the elements.  Points $A$, $B$, $C_j$, $D_j$, and $E_j$ ($j=1,2$) correspond to Fig.~\ref{fig:SimplysupDiscGRAPH} on the right. See also the video file in the electronic supplementary material.
}
    \label{fig:SimplysupDiscConfODD}
\end{figure}

The nonlinear deformation of discrete chains, in a simply supported configuration, is obtained through numerical solution of the algebraic system of equilibrium equations \eqref{eq:DiscApplicationExactEq}, supplemented by the pinned boundary conditions. In the numerical procedure, the dimensionless load $p<0$ is progressively decreased by dividing the 
load in steps $\Delta p < 0$ (of amplitude varying with the incremental stiffness) and the 
nonlinear equilibrium equations, Eqs.~\eqref{eq:DiscApplicationExactEq}, are solved using the function \lq \texttt{FindRoot}' of Wolfram Mathematica \cite{Wolfram}.

At each step of the numerical solution, the stability of the configuration is evaluated. To this purpose, being the mechanical system conservative,  the Hessian of the total potential energy can be 
obtained through differentiation of the left hand side of Eqs.~(\ref{eq:DiscApplicationExactEq}), with respect to the configuration parameters $\theta^i$ and $\beta^i$, with $i = 1, \ldots, n$. However, only $2n-1$ parameters are independent variables, so that 
the Hessian matrix must be condensed taking into account the constraint,   \cref{eq:DiscSimplySuppKinematicBoundary}, expressed in variational form as
\begin{equation}
    \sum_{i=1}^n \squarec{\roundc{\roundb{1-\alpha} \cos{\theta^i} + \alpha \cos\roundb{\theta^i + \beta^i}} \,\delta\theta^i + \alpha \cos\roundb{\theta^i + \beta^i} \,\delta\beta^i}= 0 \,.
\end{equation}

Positivity (or negativity) of the smallest eigenvalue of the condensed Hessian denotes stability (instability), while vanishing corresponds to a 
critical condition (a bifurcation or a maximum load). At bifurcation, non-trivial branches of the equilibrium path can be explored by perturbing the configuration with a small superimposed displacement parallel to the eigenvector of the Hessian corresponding to the vanishing eigenvalue.

Postcritical equilibrium paths (force $-P$ vs kinematic parameters $\theta^1$, $\beta^1$, $u_1$, and $u_2$) are reported in Fig.~\ref{fig:SimplysupDiscGRAPH} for a chain with an even (left, $n = 20$) and a odd (right, $n = 21$) number of elements.
Stable (unstable) paths are denoted in the figure with continuous (dashed) lines and small circles indicate bifurcation points. 
The two figures show similar curves both qualitatively and quantitatively, as it can be expected, by considering that two chains differ only in one element and thus must behave similarly for obvious reasons. The curves are also very similar to the response of the continuous system, Fig.~\ref{fig:SimplysupContPath}, another proof of the validity of the homogenization scheme. 

Bifurcation always occurs in a global mode. 
Postcritical deformed shapes are reported in Figs.~\ref{fig:SimplysupDiscConfEVEN} and \ref{fig:SimplysupDiscConfODD} for two chains, one with an even ($n=20$) and the other with an odd ($n=21$) number of elements, respectively.
The configuration relative to the even system are superimposed to the solution of the equivalent rod, showing a strict correspondence of the rod axis with the interpolation of the discrete chain nodes.

For both systems, a secondary bifurcation occurs when the two supports coincide, points $B$ for $n=20$ and $D_j$ ($j=1,2$) for $n=21$. 
However, the even system displays a simpler behaviour than the odd. In particular, for even number of elements, a symmetric primary bifurcation path develops after the initial bifurcation and remains stable 
up to the point where the two supports coincide, while subsequent  deformed shapes merely represent unstable configurations, a situation similar to the equivalent rod (and also to the Euler's elastica). In this system, folding corresponds to the growth of the angle formed between the two central elements of the chain. It starts to develop from the first bifurcation and continues to increase during all the postcritical path. A peculiarity of the even system is that a restabilization is achieved in the late postcritical path, indicated with point $D$ in Fig.~\ref{fig:SimplysupDiscGRAPH} on the left. 

In the odd system, the primary postcritical path initiates at bifurcation from the straight configuration and remains stable, but 
only up to a point close to but preceding the transition  load, denoted with $B$ in Fig.~\ref{fig:SimplysupDiscGRAPH} on the right. Before  that point is reached,  folding does not initiate, because the primary bifurcation satisfies symmetry and thus the central angle remains null. At point $B$, a secondary bifurcation path initiates which is stable and unsymmetric. 
The stability of the secondary bifurcation path is in line with an observation of Domokos \cite{Domokos2002} in the simpler model of a chain equivalent to the Euler's elastica. 
During the secondary path, a strong distortion of the elements develops, which is the discrete counterpart of folding in the equivalent rod. Finally, the secondary bifurcation path terminates with a tertiary bifurcation, occurring when the two supports coincide. From that point on, all equilibrium paths become unstable. 

Although only chains with $n=20$ and $21$ elements are reported, we have analysed several chains with different numbers of elements. The behaviour of these structures was always found similar to each other, showing the secondary and, in the odd case, tertiary bifurcation. The mechanics of the discrete systems was found to be perfectly consistent with the mechanics of the equivalent rod (see the video files with the progressive deformation of the continuous and discrete structures in the electronic supplementary material), thus confirming the effectiveness of the homogenization scheme and the formation and growth of folding. 

\subsection{Multiple folding}

\begin{figure}[!h]
     \centering
        \includegraphics[width=120mm]{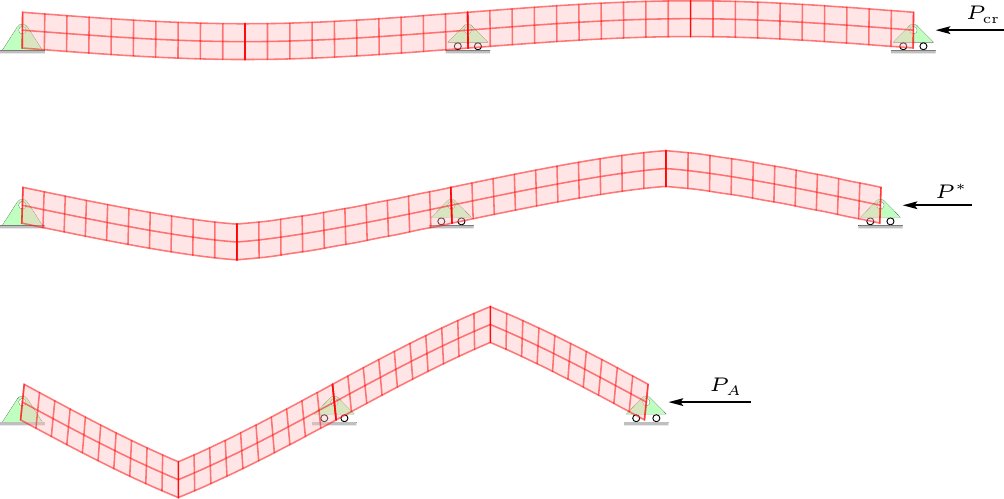}
        \caption{
        Postcritical deformed configurations of a two-span continuous rod showing the development of multiple folding under an increasing compressive load $P=\{P\ped{cr}\,,P^*\,,P_A\}$. 
    Three loading stages are displayed: (upper part) first critical load $P\ped{cr
    }$; (central part)  buckling load for the four-bar mechanism $P^*$, for which 
    folding emerges at two cross sections along  the rod axis; and (lower part) postcritical load $P_A$ at which folding is fully developed.
    }
    \label{fig:MultiFold}
\end{figure}

Folding is influenced by several factors, including structural characteristics, static and kinematic  boundary conditions, as well as the magnitude of the deformation, and  can occur at more than one cross section along the rod axis. 
Leaving to future work a thorough analysis of this peculiar response, a basic example is anticipated in
Fig.~\ref{fig:MultiFold} where the  postcritical deformed configuration is  reported for a rod of total length $L$, $\alpha=0.3$ and stiffness ratio $\zeta=20$, supported with three equally-spaced rollers under three levels of increasing compressive load $P$, showing the simultaneous folding at two different locations along the rod axis.

The treatment of the folding was performed with the technique of adding elastic hinges of increasing stiffness, as explained in the previous Section.
The deformed shape reported in the upper part of the figure refers to the (linearized) critical mode at  bifurcation, with the critical load $P\ped{cr} = -5.71 EI/L^2$ higher in magnitude than the one of the simply supported scheme ($P\ped{cr} = -4.71 EI/L^2$). The other two deformations refer to two subsequent loading steps during the postcritical behaviour, the central part to the load $P^* = -6  EI/L^2$ at which folding occurs, while the lower to a later stage, corresponding to 
the same load $P_A = - 9.02  EI/L^2$ considered in Fig.~\ref{fig:ContDeformFunctions}.

Compared with the deformed shapes reported in Figs.~\ref{fig:ContDeformFunctions}, 
\ref{fig:SimplysupContPath}, 
and 
\ref{fig:SimplysupDiscConfEVEN}, 
now the 
folding becomes more pronounced, as the parts of the elastic rod comprised between folding points remain less deformed, so that they appear to be straight. This effect is due to the fact that now the buckling load is higher than that pertaining to the simply supported scheme, a circumstance magnifying folding. In fact, on one hand $P\ped{cr}$ is closer to the load $P^*$ and the  folding occurs earlier in the post-buckling behaviour. On the other hand, recalling Eqs.~\eqref{eq:ContSimplySuppBifurcationShapes}, the relative contribution of the shear angle $\gamma$ with respect to the rotation $\theta$ at bifurcation is increased by the higher (negative) critical load, making the rod axis deformed primarily by shear effects.

\section{Conclusions}

Through homogenization, the design of microstructured chains allows to obtain {\it different} models of equivalent continuous rods, implemented with shear and bending deformability, which can be reduced as special cases to the  {\it nonlinear} models of Euler, Reissner, and Engesser.  In this vein, a new continuous model of shearable elastic rod has been introduced, which represents  the homogenized response of a discrete origami-like chain, based on  four-bar linkages and deformed in the large deformation (nonlinear) range. The derived shearable rod model reduces as particular cases to the Euler and Engesser (but not Reissner) elasticae, and was shown to be governed by a set of differential-algebraic equations. Differently from both Euler and Reissner rods, the presented model is capable 
of displaying folding in the postcritical response, even occurring at  multiple sections.
 Moreover, the equivalent rod 
shows bifurcation modes evidencing faulting deformations, involving jumps in displacement.
These mechanical features, useful in the realization of origami-based soft robot arms, have been all substantiated through the analysis of the behaviour of the underlying discrete model, where they correspond to secondary bifurcations and instabilities. 

\paragraph*{Dedication.} This article is dedicated to Professor Pedro Ponte Casta\~neda. DB values years of friendship, dedication to science, and openness to sharing ideas in the field of mechanics. His contributions have been deeply appreciated.

\paragraph*{CRediT authorship contribution statement.}
M. Paradiso: Conceptualization; Formal analysis; Investigation; Methodology; Software; Data curation; Validation;
Visualization; Writing - original draft; Writing - review and editing. 
F. Dal Corso: Conceptualization; Formal analysis; Investigation; Methodology; Supervision; Writing - review and editing.
D. Bigoni: Conceptualization; Project administration; Formal analysis; Investigation; Methodology; Supervision; Writing - original draft; Writing - review and editing; funding acquisition. 

\paragraph*{Acknowledgments.} All the authors acknowledge funding from the European Research Council (ERC) under the European Union's Horizon Europe research and innovation programme, Grant agreement No. ERC-ADG-2021-101052956-BEYOND.

\paragraph*{Disclaimer.} Views and opinions expressed in this article are those of the authors only and do not necessarily reflect those of the European Union or the ERC Executive Agency. Neither the European Union nor the granting authority can be held responsible for them.

\section*{References}
\printbibliography[heading=none]

\newpage

\begin{appendices}

\section{Matrix Eigenvalues}

Equations \eqref{eq:EigenValuesFF} and \eqref{eq:EigenValuesHH}, instrumental to
explicitly evaluate the eigenvalues of the matrix $\BF_k$, \cref{eq:EigenMatrixFF}, and $\BH_k$, \cref{eq:EigenMatrixHH}, are derived here by adapting results reported in \cite[Chapter~3]{Pal.1991}.

\subsection{Eigenvalues of the symmetric matrix \texorpdfstring{$\BF_k$}{Fk}}\label{sec:MatrixEigenFF}

Denote by $P_k(\lambda) = \det\roundb{\lambda \BI_k - \BF_k}$  the characteristic polynomial of the $k \times k$ matrix $\BF_k$, given by \cref{eq:EigenMatrixFF}. Through a Laplace expansion, $P_k(\lambda)$ can recursively be expressed as
\begin{equation}\label{eq:RecursionP}
\begin{aligned}
    P_0(\lambda) & = 1 \,, \\
    P_1(\lambda) & = \lambda - 1 \,, \\
    P_k(\lambda) & = \lambda \, P_{k-1}(\lambda) - P_{k-2}(\lambda) \,,
            \qquad k = 2, 3, \ldots \,.
\end{aligned}
\end{equation}
The change of variable $\lambda = 2x$, shows that $P_k(2 x) = V_k(x)$, where   $V_k(x)$ is the $k$--th degree Chebyshev polynomial of third kind defined as 
\begin{equation}
    V_k\roundb{\cos{\theta}} = \dfrac{\cos\roundb{\roundb{k + 1/2}\theta}}{\cos\roundb{\theta/2}} \,,
        \qquad k = 0,1,2, \ldots \,,
\end{equation}
so that the eigenvalues $\lambda_m$ of $\BF_k$ can be found by considering the $k$ roots of the polynomial $V_k(x)$ and can be expressed as in \cref{eq:EigenValuesFF}.

\subsection{Eigenvalues of the matrix \texorpdfstring{$\BH_k$}{Hk}}\label{sec:MatrixEigenHH}

The evaluation of the eigenvalues of the $k \times k$ matrix $\BH_k$, defined by \cref{eq:EigenMatrixHH}, requires the introduction of the auxiliary matrix
\begin{equation}
\setlength\arraycolsep{3pt}
\BG_k(\lambda)=
\left[
\phantom{\hspace{-1.4em}\begin{matrix} 0\\[-.2em] 0\\[-.4em] \ddots\\[-.4em] \ddots\\[-.4em] \ddots\\[-.4em] \ddots\\[-.1em] 0 \end{matrix}}
\right.
    \underbrace{
    \begin{matrix}
        1 & 1 & 1 & \cdots &\cdots & \cdots & 1 \\[-.2em]
        -1 & \lambda & -1 & 0 & \cdots & \cdots& 0 \\[-.4em]
        0 & -1 & \lambda & \ddots & \ddots & & \vdots \\[-.4em]
        \vdots & \ddots & \ddots& \ddots  & \ddots & \ddots & \vdots \\[-.4em]
        \vdots &  & \ddots & \ddots & \ddots & -1 & 0 \\[-.4em]
        \vdots & & &\ddots & -1 & \lambda & -1 \\[-.1em]
        0 & \cdots & \cdots & \cdots & 0 & -1 & \roundb{\lambda-1}
    \end{matrix}
    }_{k}
\left.
\phantom{\hspace{-1.4em}\begin{matrix} 0\\[-.2em] 0\\[-.4em] \ddots\\[-.4em] \ddots\\[-.4em] \ddots\\[-.4em] \ddots\\[-.1em] 0 \end{matrix}}
\right]
\left.
\phantom{\hspace{-1.4em}\begin{matrix} 0\\[-.2em] 0\\[-.4em] \ddots\\[-.4em] \ddots\\[-.4em] \ddots\\[-.4em] \ddots\\[-.1em] 0 \end{matrix}}
\right\}{\scriptstyle{k}}
\,,
\end{equation}
whose determinant, evaluated by means of a Laplace expansion, yields the polynomial $Q_k(\lambda)$ in the recursive form
\begin{equation}\label{eq:RecursionQ}
\begin{aligned}
    Q_1(\lambda) & = 1 \,, \\
    Q_2(\lambda) & = \lambda \,, \\
    Q_k(\lambda) & =  P_{k-1}(\lambda) + Q_{k-1}(\lambda) \,,
            \qquad k = 3, 4, \ldots \,,
\end{aligned}
\end{equation}
where $P_{k-1}(\lambda)$ is the polynomial expressed via  Eq.~(\ref{eq:RecursionP}).

Note that $Q_k(\lambda)$ coincides with the Chebyshev polynomial of second kind of degree $k-1$, $U_{k-1}(\lambda/2)$. In fact, the recursive properties evidenced in \cref{eq:RecursionP,eq:RecursionQ} lead by induction to 
\begin{equation}
    Q_k(\lambda) = \lambda \, Q_{k-1}(\lambda) - Q_{k-2}(\lambda) \,,
        \qquad k = 3, 4, \ldots \,,
\end{equation}
so that, setting $\lambda = 2x$ and considering the starting values in \cref{eq:RecursionQ}, one easily finds
\begin{equation}\label{eq:RecursionChebyQ}
\begin{aligned}
    Q_1(2x) & = U_0(x) =  1 \,, \\
    Q_2(2x) & = U_1(x) = 2x \,, \\
    Q_{k+1}(2x) & = U_k(x) = 2x \, U_{k-1}(x) - U_{k-2}(x) \,,
        \qquad k = 2, 3, \ldots \,.
\end{aligned}
\end{equation}

In view of the above preliminary result,  the eigenvalues of the matrix $\BH_k$,  \cref{eq:EigenMatrixHH}, can be evaluated by considering the characteristic polynomial $D_k(\lambda) = \det\roundb{\lambda \BI_k - \BH_k}$. A Laplace expansion leads to 
\begin{equation}
    \begin{aligned}
        D_0(\lambda) & = 1 \,, \\
        D_1(\lambda) & = \lambda \,, \\
        D_k(\lambda) &= \roundb{\lambda +1} P_{k-1}(\lambda) + Q_{k-2}(\lambda) \,, 
                \qquad k = 3, 4, \ldots \,,
    \end{aligned}
\end{equation}
which, exploiting the recursive identities in \cref{eq:RecursionP,eq:RecursionQ}, satisfies
\begin{equation}
    D_k(\lambda) = Q_{k+1}(\lambda) \,, \qquad k = 0,1,2, \ldots \,.
\end{equation}

Consequently, the eigenvalues of the matrix $\BH_k$ are given by the $k$ roots of the polynomial $ Q_{k+1}(\lambda)$ which, recalling \cref{eq:RecursionChebyQ}, can be evaluated using the properties of the Chebychev polynomials of second kind. Specifically, setting $\lambda = 2 \cos\theta$ and using the identity
\begin{equation}
    U_k(\cos\theta) = \dfrac{\sin\roundc{\roundb{k+1}\theta}}{\sin\theta} \,,
        \qquad k = 0,1,2, \ldots \,,
\end{equation}
expression \eqref{eq:EigenValuesHH} of the eigenvalues of $\BH_k$ follows.

\end{appendices}

\end{document}

%% file: Preamble.tex
\usepackage[latin1]{inputenc}
\usepackage[english]{babel}
\usepackage{mathtools,nccmath,multirow,arydshln}
\usepackage{empheq}
\usepackage{amsthm}
\usepackage{amssymb}
\usepackage{upgreek}
\usepackage{import,color}
\usepackage{xcolor}
\usepackage{graphicx}
\usepackage{subcaption}
\usepackage[title]{appendix}

\usepackage[affil-it,auth-sc]{authblk}

\usepackage{csquotes}

\usepackage[colorlinks=true,allcolors=blue]{hyperref}
\usepackage[capitalise]{cleveref} 



%% file: DefQuantities.tex
\newcommand{\Bnull}{\boldsymbol{0}}

\newcommand{\Bunitary}{\boldsymbol{1}}

\newcommand{\ds}{\mathrm{d}s}

\newcommand{\dsigma}{\mathrm{d}\sigma}

\newcommand{\dxi}{\mathrm{d}\xi}

